\documentclass[12pt]{article}

\usepackage{amsmath}
\usepackage{amssymb}
\usepackage[mathscr]{eucal}
\usepackage{graphicx}

\newtheorem{Theorem}{\sc Theorem}%%%[section]

\newtheorem{Lemma}[Theorem]{\sc Lemma}

\newtheorem{Example}[Theorem]{\sc Example}

\def\sqr#1#2{{
    \vcenter{
         \vbox{\hrule height.#2pt
               \hbox{\vrule width.#2pt height#1pt \kern#1pt
                     \vrule width.#2pt
               }
               \hrule height.#2pt
         }
    }
}}
\def\square{\mathchoice\sqr84\sqr84\sqr{2.1}3\sqr{1.5}3}

\def\essinf{\mathop{\rm ess\, inf}}
\def\esssup{\mathop{\rm ess\, sup}}

\def\bar{\overline}
\def\real{\mathbb{R}}

\newcommand{\bu}{\mbox{\boldmath{$u$}}}
\newcommand{\bv}{\mbox{\boldmath{$v$}}}

\newcommand{\bx}{\mbox{\boldmath{$x$}}}

\newcommand{\fb}{\mbox{\boldmath{$f$}}}

\newcommand{\bxi}{\mbox{\boldmath{$\xi$}}}

\newcommand{\bsigma}{\mbox{\boldmath{$\sigma$}}}
\newcommand{\btau}{\mbox{\boldmath{$\tau$}}}
\newcommand{\bvarepsilon}{\mbox{\boldmath{$\varepsilon$}}}
\newcommand{\bnu}{\mbox{\boldmath{$\nu$}}}

\newcommand{\bzeta}{\mbox{\boldmath{$\zeta$}}}

\newcommand{\bq}{\mbox{\boldmath{$q$}}}

\newcommand{\R}{{\if mm {\rm I}\mkern -3mu{\rm R}\else \leavevmode
\hbox{I}\kern -.17em\hbox{R} \fi}}

\def\lista#1
{{ \itemindent 0.0cm \labelsep .2cm \leftmargin 0.8cm \rightmargin
0.0cm \labelwidth 0.6cm \topsep 0.0mm
\parsep 0.0mm
\itemsep 0.0mm
\begin{list}{}
{ \setlength{\leftmargin}{1.0cm} \setlength{\rightmargin}{0.0cm}
\setlength{\parsep}{.0mm} \setlength{\topsep}{.0mm}
\setlength{\parskip}{.0cm} \setlength{\itemsep}{.0cm} }
{#1}\end{list}} }

\begin{document}

\title{ 
A class of dynamic frictional contact problems governed by 
a system of hemivariational inequalities in thermoviscoelasticity 
\thanks{\ This research was supported by the Marie Curie International Research Staff
Exchange Scheme Fellowship within the 7th European Community
Framework Programme under Grant Agreement No. 295118. The first author 
is also partially supported by the National Science Center of Poland 
under grant no. N N201 604640. 
}  \  }

\author{
Stanis{\l}aw Mig\'orski \thanks{\ 
The corresponding author. Email: stanislaw.migorski@ii.uj.edu.pl} \  
and Pawe{\l} Szafraniec \\
Jagiellonian University \\  
Institute of Computer Science \\ 
Faculty of Mathematics and Computer Science \\
ul. {\L}ojasiewicza 6, 30348 Krakow, Poland}

\date{}

\maketitle
\thispagestyle{empty}

\begin{center}
{\it Dedicated to the memory of Professor Zdzislaw Naniewicz}
\end{center} 

\

\noindent {\bf Abstract.} \ 
In this paper we prove the existence and uniqueness of the weak 
solution for a dynamic thermoviscoelastic problem which describes 
frictional contact between a body and a foundation. We employ 
the nonlinear constitutive viscoelastic law with a long-term memory, 
which include the thermal effects and consider 
the general nonmonotone and multivalued subdifferential boundary conditions
for the contact, friction and heat flux. The model consists of the system of 
the hemivariational inequality of hyperbolic type for the displacement 
and the parabolic hemivariational inequality for the temperature. 
The existence of solutions is proved by using recent results 
from the theory of hemivariational inequalities and a fixed point argument. 

\vskip 4mm

\noindent {\bf Keywords:} Dynamic contact; thermoviscoelastic; 
evolution hemivariational inequality; Clarke subdifferential; 
nonconvex; hyperbolic; parabolic; viscoelastic material; 
frictional contact; weak solution.

\vskip 4mm

\noindent {\bf Mathematics Subject Classification 2000: } 
74M15, 35L70, 74H20, 35L85, 49J40.

\newpage

%%%%%%%%%%%%%%%%%%%%%%%%%%%%%%%%%%%
\section{Introduction}\label{Introduction}
%%%%%%%%%%%%%%%%%%%%%%%%%%%%%%%%%%%

\noindent 
Problems involving thermoviscoelastic contact arise naturally in many situations, 
particularly those involving industrial processes when two or more deformable 
bodies may come in contact or may lose contact as a result of thermoviscoelastic
expansion or contraction. For this reason there is a considerable literature devoted 
to this topic. The first existence and uniqueness results for contact problems with friction 
in elastodynamics were obtained by Duvaut and Lions \cite{DL}. Later, 
Martins and Oden \cite{MO} studied the normal compliance model 
of contact with friction and showed existence and uniqueness results 
for a viscoelastic material. These results were extended by Figueiredo 
and Trabucho \cite{FT} to thermoelastic and thermoviscoelastic models. 
In these papers the authors used the classical Galerkin method 
combined with a regularization technique and compactness arguments. 
Recently dynamic viscoelastic frictional contact problems with or without 
thermal effects have been investigated in a large number of papers, 
see e.g. 
Adly et al.~\cite{ADLY}
Amassad et al. \cite{AKRS}, 
Andrews et al. \cite{A2002, A1997}, 
Chau et al. \cite{CHAU2}, 
Han and Sofonea \cite{HAN}, 
Jarusek \cite{JAR}, 
Kuttler and Shillor \cite{KS2001}, 
Migorski \cite{MIGORSKI7}, 
Migorski and Ochal \cite{MIGORSKIOPT}, 
Migorski et al.~\cite{MOS1,MOSBOOK},
Rochdi and Shillor \cite{RS2000} 
and the references therein. 

In this paper we consider the frictional contact problem between a nonlinear 
thermoviscoelastic body and an obstacle. We suppose that the process 
is dynamic and the material is viscoelastic with long memory and thermal 
effect. Our main interest lies in general nonmonotone and possibly multivalued 
subdifferential boundary conditions. More precisely, it is supposed 
that on the contact part of the boundary of the body under consideration, 
the subdifferential relations hold, 
the first one between the normal component of the velocity 
and the normal component of the stress, 
the second one between the tangential components of these quantities 
and the third one between temperature and the heat flux vector. 
These three subdifferential boundary conditions are the natural 
generalizations of the normal damped response condition, 
the associated friction law and the well known Fourier law 
of heat conduction, respectively. 
For examples, applications and detailed explanations concerning 
the boundary conditions we refer to Panagiotopoulos~\cite{PANA1, PANA7}, 
Naniewicz and Panagiotopoulos~\cite{NP}, and Migorski et al.~\cite{MOSBOOK}. 

The thermoviscoelastic phenomena can be divided into three classes: 
static, quasistatic, and full dynamic. 
The quasistatic problems can be viewed as being of mixed elliptic--parabolic 
type, while the dynamic case is of mixed hyperbolic--parabolic type. 
The latter is more complicated, and we have in the literature only a few results 
concerning existence and uniqueness. 
We investigate a fully dynamic contact problem which consists 
of the energy-elasticity equations of hyperbolic type together 
with a nonlinear parabolic equation for the temperature. 
Because of the multivalued multidimensional boundary conditions, 
the problem is formulated as a system of two coupled evolution 
hemivariational inequalities. 
All subdifferentials are understood in this paper in the sense of Clarke 
and are considered for locally Lipschitz, and in general nonconvex 
and nonsmooth superpotentials. 
This allows to incorporate in our model several types of boundary 
conditions considered earlier e.g. in 
\cite{NP, PANA1, PANA7, MOSBOOK}. 
We note that when the superpotentials involved in the problem are convex 
functions, the hemivariational inequalities reduces to variational inequalities. 

\

. . . . . . . . . 

\

The goal of the paper is to provide the result on existence and 
uniqueness of a global weak solution to the system. 
The existence of solutions is obtained by combining 
recent results on the hyperbolic hemivariational 
inequalities~\cite{MIGORSKI3,MIGORSKI6,MOSBOOK,Kulig2,Kulig1} 
and the results on the parabolic hemivariational 
inequalities~\cite{MIGORSKI2,MIGORSKI8}, 
and by applying a fixed point argument. 
In spite of importance of the subject in applications, 
to the best of the authors' knowledge, the existence of solutions 
to the system of hemivariational inequalities in dynamic 
thermoviscoelasticity has studied in very 
few papers~\cite{DMTHERMO,Denkowski,DMO}
However, in all aformentioned papers, there is a coupling 
between the displacement (and velocity) and the temperature 
in the constitutive law which is assumed to be linear. 
In this paper we deal with the fully nonlinear constitutive relation 
and assume the coupling also in the heat flux boundary condition 
on the contact surface.
Finally, we note that for linear thermoelastic materials a system of 
hemivariational inequalities was formulated by Panagiotopoulos 
in Chapter 7.3 of~\cite{PANA7}. However, the regularity hypotheses 
on the multivalued terms were quite unnatural and the data 
were assumed to be very regular (cf. Proposition~7.3.2 in~\cite{PANA7}).

The content of the paper is as follows. After the preliminary material 
of Section~\ref{Preliminaries}, in Section~\ref{Physical} we present the physical 
setting and the classical formulation of the problem. 
In Section~\ref{Weak} we deliver the variational formulation 
of the mechanical problem and state our main existence and uniqueness result. 
The proof of the main result is provided in Section~\ref{Main}.  
Some examples of nonmonotone and multivalued subdifferential 
boundary conditions are given in Section~\ref{Examples}.

\section{Preliminaries}\label{Preliminaries}
%%%%%%%%%%%%%%%%%%%%%

\noindent 
In this section we introduce notation and recall some definitions 
and results needed in the sequel, cf. \cite{HAN,DMP2, MOSBOOK,NOWACKI,PANA1}.

We denote by $\mathbb{S}^{d}$ the linear space of second order symmetric 
tensors on $\real^d$, $d = 2$, $3$, or equivalently, the space 
$\real_s^{\, d \times d}$ of symmetric matrices of order $d$. 
We recall that the canonical
inner products and the corresponding norms on $\mathbb{R}^d$ and
$\mathbb{S}^{d}$ are given 
by
\begin{align*}
\bu\cdot\bv & =  u_i \, v_i, \quad
  \|\bv\|_{\real^d} =(\bv\cdot\bv)^{1/2}\quad \mbox{for all} \ \
  \bu=(u_i),\ \bv=(v_i)\in \mathbb{R}^d, \\
\bsigma : \btau & =\sigma_{ij}\, \tau_{ij},\quad
  \|\btau\|_{\mathbb{S}^{d}}
  =(\btau : \btau)^{1/2}\quad \mbox{for all} \quad \bsigma=(\sigma_{ij}),\,
  \btau=(\tau_{ij}) \in\mathbb{S}^{d},
\end{align*}

\noindent 
respectively.  
Here and below, the indices $i$ and $j$ run from $1$ to $d$, and  
the summation convention over repeated indices is adopted.

Let $\Omega$ be an open bounded subset of $\real^d$ with a Lipschitz 
continuous boundary $\Gamma$ and let $\bnu$ denote the outward unit normal 
vector to $\Gamma$. 
We introduce the spaces
\begin{equation*}
H = L^2(\Omega; \mathbb{R}^{d}), \
{\mathcal H} = \left\{
\btau = (\tau_{ij}) \, \mid \, \tau_{ij} = \tau_{ji} \in L^2(\Omega) \right\}, \ 
{\mathcal H}_1 = \left\{\, \btau \in {\mathcal H} \, \mid \, {\rm Div}\, \btau \in H \,\right\}.
\end{equation*}

\noindent 
It is well known that the spaces $H$, ${\mathcal H}$ and ${\mathcal H}_1$
are Hilbert spaces equipped with the inner products
\begin{equation*}
\langle \bu, \bv \rangle_{H} = \int_\Omega \bu \cdot \bv \, dx, \
\langle \bsigma, \btau \rangle_{\mathcal H} = \int_\Omega
\bsigma \, \colon \, \btau \, dx, \ 
\langle \bsigma, \btau \rangle_{{\mathcal H}_1} = \langle
\bsigma, \btau \rangle_{\mathcal H} + \langle {\rm Div}\,\bsigma,
{\rm Div}\,\btau \rangle_{H},
\end{equation*}

\noindent where $\bvarepsilon \colon H^1(\Omega;\real^d) \to
{\mathcal H}$ and ${\rm Div} \colon {\mathcal H}_1 \to H$ denote
the deformation and the divergence operator, respectively,
given by
\begin{equation*}%%\label{def-div}
\bvarepsilon(\bu) = \big(\varepsilon_{ij} (\bu) \big), \ \ \
\varepsilon_{ij}(\bu) = \frac{1}{2} \, (u_{i,j} + u_{j,i}), \ \ \
{\rm Div} \, \bsigma = (\sigma_{ij,j}).
\end{equation*}

\noindent 
An index that follows a comma indicates a derivative with respect to 
the corresponding component of the spatial variable $\bx \in \Omega$. 
Given $\bv \in H^1(\Omega; \real^d)$ we denote by $\gamma_0 \bv$ its trace on $\Gamma$, 
where $\gamma_0 \colon H^1(\Omega;\real^{\, d}) \to 
H^{1/2}(\Gamma; \real^{\, d}) \subset L^{2}(\Gamma; \real^{\, d})$ 
is the trace map. 
If $d = 1$, then the trace operator from $H^1(\Omega)$ into $L^2(\Gamma)$ 
is denoted by $\gamma_0^s$. 
For $\bv \in L^2(\Gamma; \real^{\, d})$ 
we denote by $v_\nu$ and $\bv_\tau$ the usual normal and tangential 
components of $\bv$ on the boundary $\Gamma$, i.e.,  
$v_\nu = \bv \cdot \bnu$ and $\bv_\tau = \bv - v_\nu \bnu$. 
Similarily, for a regular tensor field $\bsigma \colon \Omega \to \mathbb{S}^{d}$, 
we define its normal and tangential components by 
$\sigma_\nu = (\bsigma \bnu) \cdot \bnu$ and 
$\bsigma_\tau = \bsigma \bnu - \sigma_\nu \bnu$, respectively. 
The following two Green--type formulas can be found in 
Chapter~2 of \cite{MOSBOOK}:
\begin{equation}\label{green1}
\int_\Omega \left( u \, {\rm div} \, \bv + \nabla u\cdot \bv
\right) \, dx = \int_\Gamma u \, (\bv \cdot \bnu) \, d\Gamma,
\end{equation}

\noindent
for $u \in H^1(\Omega)$ and $\bv \in H^1(\Omega;\real^d)$, 
and 
\begin{equation}\label{green2}
\int_\Omega \bsigma:\bvarepsilon (\bv)\,dx
+ \int_\Omega {\rm Div}\,\bsigma \cdot \bv\,dx
= \int_\Gamma \bsigma\bnu\cdot\bv\,d\Gamma
\end{equation}

\noindent for $\bv \in H^1(\Omega;\real^d)$ and $\bsigma \in
C^1({\bar{\Omega}; {\mathbb{S}^d}})$.

We recall the definitions of the generalized directional 
derivative and the generalized gradient of Clarke for a locally
Lipschitz function $\varphi \colon X \to \real$, where $X$ is a Banach
space (see \cite{CLARKE}). 
The generalized directional derivative of $\varphi$ at $x \in X$ in the
direction $v \in X$, denoted by $\varphi^{0}(x; v)$, is defined by
$$\displaystyle 
\varphi^{0}(x; v) =
\limsup_{y \to x, \ t\downarrow 0}
\frac{\varphi(y+tv) - \varphi(y)}{t}.
$$
The generalized gradient of $\varphi$ at $x$, denoted by
$\partial \varphi(x)$, is a subset of a dual space $X^*$ given by
$\partial \varphi(x) = \{ \zeta \in X^* \mid \varphi^{0}(x; v) \ge 
{\langle \zeta, v \rangle}_{X^* \times X}$  
for all $v \in X \}$. 
 
We denote by ${\mathcal L}(X, Y)$ the space of linear continuous 
mappings from $X$ to $Y$. 
Given a reflexive Banach space $Y$, we denote by
${\langle \cdot, \cdot \rangle}_{Y^* \times Y}$ 
the duality pairing between the dual space $Y^*$ and $Y$. 
In what follows different positive constants, which may change 
from line to line, will be denoted by the same letter $c$.

Finally, we recall the following result (cf. Lemma 7 in \cite{Kulig1}) 
which is a consequence of the Banach contraction principle 
and which will be used in the proof of the main theorem of this paper.  
\begin{Lemma}\label{LemmaKulig}
Let $X$ be a Banach space with a norm $\| \cdot \|_X$ and $T > 0$. 
Let $\Lambda \colon L^2(0,T; X) \to L^2(0,T; X)$ be an operator satisfying 
\begin{equation*}
\| (\Lambda \eta_1)(t) - (\Lambda \eta_2)(t)\|^2_X \le 
c \int_0^t \| \eta_1(s) - \eta_2(s)\|^2_X \, ds 
\end{equation*}

\noindent 
for every $\eta_1$, $\eta_2 \in L^2(0,T;X)$, a.e. $t \in (0,T)$ with a constant $c > 0$. 
Then $\Lambda$ has a unique fixed point in $L^2(0, T; X)$, 
i.e. there exists a unique $\eta^* \in  L^2(0,T;X)$ such that $\Lambda \eta^* = \eta^*$.
\end{Lemma}

%%%%%%%%%%%%%%%%%%%%%%%%%%%%%%%%
\section{Physical setting and classical formulation}\label{Physical}
%%%%%%%%%%%%%%%%%%%%%%%%%%%%%%%%

\noindent 
In this section we introduce the physical setting of the problem, 
describe the classical model and list the hypotheses on the data.

Let $\Omega$ be an open bounded domain in $\real^d$, $d = 2$, $3$, 
with a Lipschitz continuous boundary $\Gamma = \partial \Omega$. 
The boundary $\Gamma$ is composed of three 
sets $\overline{\Gamma}_D$, $\overline{\Gamma}_N$ and
$\overline{\Gamma}_C$, with mutually disjoint relatively open sets
$\Gamma_D$, $\Gamma_N$ and $\Gamma_C$, such that 
${\rm meas} \, (\Gamma_D) >0$.  
We consider a viscoelastic body, which in the reference configuration, 
occupies volume $\Omega$ and which is supposed to be stress free 
and at a constant temperature, conveniently set as zero. 
We assume that the temperature changes accompanying 
the deformations are small and they do not produce any changes 
in the material parameters which are regarded temperature independent. 
We are interested in a mathematical model that describes the evolution 
of the mechanical state of the body and its temperature during 
the time interval $[0, T]$ where $0< T < \infty$. 
To this end, we denote by 
$\bsigma=\bsigma(\bx,t)=(\sigma_{ij}(\bx,t))$ the stress field, 
by
$\bu=\bu(\bx,t)=(u_i(\bx,t))$ the displacement field, and 
by
$\theta = \theta (\bx, t)$ the temperature, 
where $\bx \in \Omega$ and $t \in [0, T]$ 
denote the spatial and the time variables, respectively. 
The functions 
$\bu \colon {\Omega}\times[0,T] \to \mathbb{R}^d$, 
$\bsigma \colon {\Omega}\times[0,T] \to \mathbb{S}^d$ and 
$\theta \colon \Omega \times [0, T] \to \real$ will play the role of
the unknowns of the frictional contact problem. 
From time to time, we suppress the explicit dependence 
of the quantities on the spatial variable $\bx$, or both $\bx$ and $t$. 

We suppose that the body is clamped on $\Gamma_D$, 
the volume forces of density $f_0 = f_0(x, t)$ act in $\Omega$ and 
the surface tractions of density $f_1 = f_1(x, t)$ are applied on $\Gamma_N$. 
Moreover, the body is subjected to a heat source term per unit volume 
$g = g(x, t)$ and it comes in contact with an obstacle, 
the so-called foundation, over the contact surface $\Gamma_C$. 
We also use the notation 
$Q = \Omega \times (0, T)$,   
$\Sigma_D = \Gamma_D \times (0, T)$,
$\Sigma_N = \Gamma_N \times (0, T)$ 
and 
$\Sigma_C = \Gamma_C \times (0, T)$. 
Without loss of generality we can assume that the material density and 
the specific heat at constant deformation are constants, both set equal to one. 
Assuming small displacements, the system of the equation of motion and 
the law of conservation of energy take the form
\begin{eqnarray*}
u'' (t) - {\rm Div} \, \bsigma (t) = f_0(t) & \mbox{in} & Q \\ [2mm]
\theta' (t) + {\rm div} \, \bq (t) = R(t, \bu'(t)) + g(t) & \mbox{in} & Q.  
\end{eqnarray*}

\noindent 
For the thermal diffussion, we adopt the following law with 
the heat flux vector $\bq$ of the form 
$$
\bq(t) = - K(\bx,t, \nabla \theta (t)) \ \ \mbox{in} \ \ Q .
$$ 

\noindent 
In the case $K(\bx, t, \cdot)$ is a linear function, this law reduces 
to the Fourier law of heat conduction of the form 
$\bq (t) = - k(\bx, t) \nabla \theta(t)$ 
in $Q$ where $k = k(\bx, t)$ represents the thermal conductivity tensor. 
In the heat equation, we suppose that $R$ is a nonlinear function 
of the velocity. A model with a linear function $R$ of the form 
$R(x, t, v) = - \sum_{i,j=1}^d c_{ij}(x, t) \frac{\partial v_i}{\partial x_i}$ 
for $v \in H^1(\Omega; \real^d)$, 
a.e. $(\bx, t) \in Q$, where $c_{ij} \in L^\infty(Q)$ are the components of the tensor 
of thermal expansion was considered in~\cite{ADLY,CHAU2}. 
The behavior of the material is described by the nonlinear thermoviscoelastic constitutive 
law of Kelvin-Voigt type with a long-term memory of the form  
\begin{equation*}
\bsigma(t) = 
{\mathcal A}(t,\bvarepsilon(\bu'(t))) +
{\mathcal B}(t,\bvarepsilon(\bu(t))) +
\int_0^t\,{\mathcal C}(t-s) \bvarepsilon(\bu(s)) \, ds + {\mathcal C}_e (t, \theta (t)) 
\ \ \mbox{in} \ \ Q.
\end{equation*}

\noindent
We allow the viscosity operator ${\mathcal A}$, 
the elasticity operator ${\mathcal B}$, 
the relaxation operator ${\mathcal C}$ and 
the thermal expansion operator ${\mathcal C}_e$ 
to depend on the time. This law generalizes the following 
classical equation of the linear thermoviscoelasticity theory 
of the form
$$
\displaystyle 
\sigma_{ij} = a_{ijkl} \, \varepsilon_{kl}(u') + 
b_{ijkl} \, \varepsilon_{kl} (u) - c_{ij} \, \theta \ \ \mbox{in} \ \ Q,  
$$

\noindent 
where $a = (a_{ijkl})$ and $b = (b_{ijkl})$, $i$, $j$, $k$, $l = 1, \ldots, d$ 
are the viscosity and elasticity fourth order tensors, respectively, and 
$(c_{ij})$ are the so-called coefficients of thermal expansion. 

Our main interest lies in the contact and friction boundary conditions on the surface 
$\Gamma_C$. As concerns the contact condition we assume that the normal stress 
$\sigma_\nu$ and the normal velocity $u_\nu'$ satisfy the 
nonmonotone normal damped response condition of 
the form 
\begin{equation*}%%\label{Normal}
- \sigma_\nu \in \partial j_\nu(\bx, t, u_\nu') \ \ {\rm on} \ \ \Sigma_C .
\end{equation*}

\noindent 
The friction relation is given 
by  
\begin{equation*}%%\label{Tangential} 
- \bsigma_\tau \in \partial j_\tau(\bx, t, \bu_\tau') \ \ {\rm on} \ \ \Sigma_C  
\end{equation*} 

\noindent  
and describes the multivalued law between the tangential force 
$\bsigma_\tau$ on $\Gamma_C$ and the tangential velocity $\bu_\tau'$.  
Moreover, we suppose that there is heat exchange between the surface 
$\Gamma_C$ and the foundation and that the dependence between the 
heat flux vector and the boundary temperature is described by the possibly 
multivalued relation of the subdifferential type with a nonconvex potential $j$. 
Since the power that is generated by the frictional contact forces is proportional 
to the tangential velocity, we introduce the function $h_\tau$ in the following 
relation
$\bq (t) \cdot \bnu \in h_\tau (t, \| \bu_\tau' (t) \|_{\real^d}) - \partial j(\bx, t, \theta(t))$ 
on $\Sigma_C$. We rewrite it in the following 
form 
\begin{equation}\label{Temper} 
- \frac{\partial \theta}{\partial \nu_K} \in \partial j(\bx, t, \theta(t)) 
- h_\tau (\bx, t, \| \bu_\tau' (\bx, t) \|_{\real^d})
\ \ \mbox{on} \ \ \Sigma_C ,   
\end{equation} 

\noindent  
where $\frac{\partial \theta}{\partial \nu_K} = K(\bx, t, \nabla \theta (t)) \cdot \bnu$.  
In a simple case, when $h_\tau \equiv 0$ (there is no coupling between the 
temperature and the tangential velocity on $\Sigma_C$) 
and 
$j(\bx, t, r) = \frac{1}{2} \, k_e \, (r - \theta_R)^2$ for $r \in \real$, a.e. $(\bx, t) \in \Sigma_C$, 
$k_e$ being the heat exchange coefficient between the body 
and the foundation and $\theta_R$ being the temperature of the foundation, 
the condition (\ref{Temper}) reduces to the equation 
\begin{equation*}
- \frac{\partial \theta}{\partial \nu_K} = k_e \, (\theta - \theta_R) 
\ \ \mbox{on} \ \ \Sigma_C   
\end{equation*} 

\noindent 
which was studied in~\cite{ADLY,CHAU2}.
%%%%%%%%and $h_\tau$ is a given tangential function. 
As a simple tangential function $h_\tau$ in (\ref{Temper}), we may take
$$\displaystyle
h_\tau(\bx, t, r) = \lambda (\bx, t) \, r \ \ \mbox{for all} \ r \in \real_+, 
\ \mbox{a.e.} \ (\bx, t) \in \Sigma_C , 
$$

\noindent 
where $\lambda \in L^\infty(\Sigma_C)$ represents 
a time-dependent rate coefficient for the gradient of the temperature. 
Here 
$j_\nu \colon \Sigma_C \times \real \to \real$,  
$j_\tau \colon \Sigma_C \times \real^{\, d} \to \real$ 
and 
$j \colon \Sigma_C \times \real \to \real$ 
are locally Lipschitz functions in their last variables and 
$\partial j_\nu$, $\partial j_\tau$, $\partial j$ represent their Clarke subdifferentials. 
Many various possibilites of nonconvex potentials $j_\nu$, $j_\tau$, $j$ 
can be considered to model boundary conditions, see e.g.~\cite{MOSBOOK} 
for examples and applications. 
For the sake of simplicity, we assume that the temperature vanishes on
$\Gamma_D \cup \Gamma_N$, i.e. $\theta = 0$ on 
$(\Gamma_D \cup \Gamma_N) \times (0, T)$. 
Finally, we denote by $\bu_0$, $\bv_0$ and $\theta_0$ the initial displacement, 
the initial velocity and the initial temperature, respectively. 
Under these assumptions, the classical formulation of the mechanical problem 
of frictional contact for the thermoviscoelastic body is the following. 

\medskip 

\noindent 
{\bf Problem} $P$:
find a displacement field $\bu \colon Q \to \real^d$ and 
a temperature $\theta \colon Q \to \real$ 
such that 
\begin{eqnarray} \label{EQ1} 
u'' (t) - {\rm Div} \, \bsigma (t) = f_0(t) & \mbox{in} & Q \\
\nonumber 
\bsigma(t) = 
{\mathcal A}(t,\bvarepsilon(\bu'(t))) +
{\mathcal B}(t,\bvarepsilon(\bu(t))) + \qquad 
\\
\label{EQ2} 
+ \int_0^t\,{\mathcal C}(t-s) \bvarepsilon(\bu(s)) \, ds + 
{\mathcal C}_e (t, \theta (t))  & \mbox{in} & Q 
\\ 
\label{EQ3}  
\theta' (t) - {\rm div} \, K(x, t, \nabla \theta (t)) = R(t, \bu'(t)) + g(t) & \mbox{in} & Q 
\\
\label{EQ4}  
\bu (t) = 0 & \mbox{on} & \Sigma_D 
\\ 
\label{EQ5}  
\bsigma (t) \bnu = f_1 (t) &  \mbox{on} & \Sigma_N
\\ 
\label{EQ6}  
- \sigma_\nu \in \partial j_\nu(\bx, t, u'_\nu (t)), \ \  
- \bsigma_\tau \in \partial j_\tau(\bx, t, \bu_\tau' (t)) 
& \mbox{on} & \Sigma_C   
\\ 
\label{EQ7} 
- \frac{\partial \theta}{\partial \nu_K} \in \partial j(\bx, t, \theta(t)) 
- h_\tau (\bx, t, \| \bu_\tau' (\bx, t) \|_{\real^d})
& {\rm on} & \Sigma_C   
\\
\label{EQ8}   
\theta (t) = 0 & {\rm on} & (\Gamma_D \cup \Gamma_N) \times (0, T)
\\ 
\label{EQ9}  
\bu(0) = \bu_0, \ \ \bu'(0) = \bv_0, \ \ \theta(0) = \theta_0 & \mbox{in} & \Omega.
\end{eqnarray}

In order to provide the variational formulation of Problem $P$, 
we need some additional notation. 
We introduce the following spaces 
$$
\displaystyle 
E = \left\{ v \in H^1(\Omega; \real^d) \mid v = 0 \ \ {\rm on} \ \Gamma_D \right\} 
\ \ \mbox{and} \ \  
V = \left\{ \eta \in H^1(\Omega) \mid \eta = 0 \ \ {\rm on} \ \Gamma_D \cup \Gamma_N \right\}. 
$$ 

\noindent 
On $E$ we consider the inner product and the corresponding norm 
given by 
$$
\displaystyle 
(u, v)_E = \langle 
\bvarepsilon (u), \bvarepsilon (v) \rangle_{L^2(\Omega; \mathbb{S}^{d})}, \quad 
\| v \|_E = \| \bvarepsilon (v) \|_{L^2(\Omega; \mathbb{S}^{d})} \ \ {\rm for} \ u, v \in E. 
$$ 

\noindent 
From the Korn inequality 
$\| v \|_{H^1(\Omega; \real^d)} \le c \| \bvarepsilon (v) \|_{L^2(\Omega; \mathbb{S}^{d})}$ 
for $v \in E$ with $c > 0$, 
it follows that 
$\| \cdot \|_{H^1(\Omega; \real^d)}$ and 
$\| \cdot \|_{E}$ are the equivalent norms on $E$.
Let $H = L^2(\Omega; \real^d)$ and $Z = H^\delta(\Omega; \real^d)$ with a fixed 
$\delta \in (1/2, 1)$. 
Denoting by $i \colon E \to Z$ the embedding injection and 
by $\gamma \colon Z \to L^2(\Gamma; \real^d)$ the trace operator, 
for all $v \in E$, we have $\gamma_0 v = \gamma (i v)$. 
For simplicity we omit the notation of the embedding and write $\gamma_0 v = \gamma v$ 
for $v \in E$. 
Identifying $H$ with its dual, we have the following evolution fivefold 
of spaces with dense, continuous and compact embeddings
$$
E \subset Z \subset H \subset Z^* \subset E^*. 
$$

\noindent  
We also introduce the following spaces of vector valued functions 
$\displaystyle {\mathcal E} = L^{2}(0,T; E)$, 
$\displaystyle {\mathcal Z} = L^{2}(0,T; Z)$, 
$\displaystyle {\widehat{{\mathcal H}}} = L^{2}(0,T; H)$ and 
$\displaystyle {\mathbb{E}} = \{ v \in {\mathcal E} \mid v' \in {\mathcal E^*} \}$, 
where the time derivative is understood in the sense of vector 
valued distributions. 
Endowed with the norm 
$\| v \|_{{\mathbb{E}}} = \| v \|_{\mathcal E} + \| v' \|_{\mathcal E^*}$, 
the space ${\mathbb{E}}$ becomes a separable reflexive Banach space.
We have 
$$
{\mathbb{E}} \subset {\mathcal E} \subset {\mathcal Z} \subset {\widehat{{\mathcal H}}}   
\subset {\mathcal Z^*} \subset {\mathcal E^*}.
$$ 

\noindent 
with dense and continuous embeddings. 
The duality for the pair $({\mathcal E}, {\mathcal E}^*)$ is denoted
by 
$\langle w, z \rangle_{{\mathcal E}^* \times {\mathcal E}} = 
\int_0^T \langle w(s), z(s) \rangle_{E^* \times E} \, ds$. 
It is well known (see e.g. \cite{DMP2,ZEIDLER}) that the embeddings 
${\mathbb{E}} \subset C(0, T; H)$ and 
$\{ v \in {\mathcal E} \mid v' \in {\mathbb{E}} \} \subset C(0,T; E)$ are continuous 
and ${\mathbb{E}} \subset {\mathcal Z}$ is compact.

Similarly, we introduce the space $Y = H^\delta(\Omega)$ with the same $\delta \in (1/2, 1)$
and we obtain the evolution fivefold of spaces
$$
V \subset Y \subset L^2(\Omega) \subset Y^* \subset V^* 
$$

\noindent  
with dense, continuous and compact embeddings. 
Let 
$\displaystyle {\mathcal V} = L^2(0,T; V)$, 
$\displaystyle {\mathcal Y} = L^2(0,T; Y)$ 
and 
$\displaystyle {\mathcal W} = \{ \eta \in {\mathcal V} \mid \eta' \in {\mathcal V^*} \}$. 
We have
$$
{\mathcal W} \subset {\mathcal V} \subset {\mathcal Y} \subset L^2(0, T; L^2(\Omega)) 
\subset {\mathcal Y^*} \subset {\mathcal V^*},
$$ 

\noindent 
where all the embeddings are dense and continuous. We also know that 
the embeddings ${\mathcal W} \subset C(0, T; L^2(\Omega))$ and 
$\{ \eta \in {\mathcal V} \mid \eta' \in {\mathcal W} \} \subset C(0, T; V)$ are continuous and 
${\mathcal W} \subset {\mathcal Y}$ is compact. 
Furthermore, we denote by $\gamma_s \colon Y \to L^2(\Gamma)$ 
the trace operator for scalar valued functions and we write 
$\gamma_0^s v = \gamma_s v$ for $v \in V$.  

\medskip

The following assumptions on the data of Problem $P$ will be needed 
throughout the paper. We assume that the viscosity operator ${\mathcal A}$, 
the elasticity operator ${\mathcal B}$, the relaxation operator ${\mathcal C}$ 
and the thermal expansion operator ${\mathcal C}_e$ satisfy the following hypotheses.

\medskip

\noindent
$\underline{H({\mathcal A})}:$ \quad 
$\displaystyle {\mathcal A} \colon Q \times \mathbb{S}^{d} \to \mathbb{S}^{d}$ is such that 

%\smallskip

\lista{
\item[(a)]
${\mathcal A}(\cdot,\cdot, \varepsilon)$ is measurable on $Q$ 
for all $\varepsilon \in \mathbb{S}^{d}$.
\smallskip
\item[(b)]
${\mathcal A}(x, t, \cdot)$ is continuous on $\mathbb{S}^{d}$ for a.e. $(x, t) \in Q$.
\smallskip
\item[(c)]
$\displaystyle 
\| {\mathcal A}(x,t,\varepsilon) \|_{\mathbb{S}^{d}} \le
a_0(x,t) + a_1 \, \| \varepsilon \|_{\mathbb{S}^{d}}$ for all $\varepsilon \in \mathbb{S}^{d}$, 
a.e. $(x, t) \in Q$ with $a_0 \in L^2(Q)$, $a_0 \ge 0$, $a_1 > 0$.
\smallskip
\item[(d)]
$\displaystyle 
\left( {\mathcal A}(x, t, \varepsilon_1) - {\mathcal A} (x, t, \varepsilon_2) 
\right) : (\varepsilon_1 - \varepsilon_2) \ge 
m_{\mathcal A} \, \| \varepsilon_1 - \varepsilon_2 \|^2_{\mathbb{S}^{d}}$ 
for all $\varepsilon_1$, $\varepsilon_2 \in \mathbb{S}^{d}$, a.e. $(x, t) \in Q$ 
with $m_{\mathcal A} > 0$. 
\smallskip
\item[(e)]
$\displaystyle {\mathcal A}(x,t,\varepsilon) : \varepsilon \ge
\alpha_{\mathcal A} \, \| \varepsilon \|^2_{\mathbb{S}^{d}}$ 
for all $\varepsilon \in \mathbb{S}^{d}$, 
a.e. $(x,t) \in Q$ with $\alpha_{\mathcal A} > 0$. 
}

\medskip

%\begin{Remark}\label{newremark}
%It should be remarked that the growth condition $H({\mathcal A})$(iii) excludes terms 
%with power greater than one, but is satisfied within linearized viscoelasticity,
%and is satisfied by truncated operators, cf. \cite{HAN,SST}.
%\end{Remark}

\noindent
$\underline{H({\mathcal B})}:$ \quad 
$\displaystyle {\mathcal B} \colon Q \times \mathbb{S}^{d} \to \mathbb{S}^{d}$ is such that 

%\smallskip

\lista{
\item[(a)]
${\mathcal B}(\cdot,\cdot, \varepsilon)$ is measurable on $Q$ 
for all $\varepsilon \in \mathbb{S}^{d}$.
\smallskip
\item[(b)]
$\displaystyle 
\| {\mathcal B}(x,t,\varepsilon) \|_{\mathbb{S}^{d}} \le
b_0(x,t) + b_1 \, \| \varepsilon \|_{\mathbb{S}^{d}}$ 
for all $\varepsilon \in \mathbb{S}^{d}$,
a.e. $(x, t) \in Q$ with $b_0 \in L^2(Q)$, $b_0$, $b_1 \ge 0$.
\smallskip
\item[(c)]
$\displaystyle 
\| {\mathcal B}(x,t,\varepsilon_1) - {\mathcal B} (x,t,\varepsilon_2) \|_{\mathbb{S}^{d}} 
\le L_{\mathcal B} \| \varepsilon_1 - \varepsilon_2 \|_{\mathbb{S}^{d}}$ 
for all $\varepsilon_1$, $\varepsilon_2 \in \mathbb{S}^{d}$, 
a.e. $(x, t) \in Q$ with $L_{\mathcal B} > 0$. 
}

%\begin{Remark}\label{newBremark} 
%If ${\mathcal B} (x, t, \cdot) \in {\mathcal L}(\mathbb{S}^{d}, \mathbb{S}^{d})$
%for a.e. $(x, t) \in Q$, the conditions $H({\mathcal B})${\rm (ii)} and {\rm (iii)} hold. 
%Thus the hypothesis $H({\mathcal B})$ is more general than the ones considered 
%in~\cite{AA2005,JOGO2005,Guiyang,MO1,Unified,AO,PARK1} where 
%the elasticity operator is assumed to be linear (which corresponds to the Hooke law). 
%\end{Remark}

\medskip

\noindent $\underline{H({\mathcal C})}:$ \quad 
${\mathcal C} \colon Q \times \mathbb{S}^{d} \to \mathbb{S}^{d}$ is such that 

\lista{
\item[(a)]
${\mathcal C}(x,t,\varepsilon) = c(x,t) \, \varepsilon$ 
for all $\varepsilon \in \mathbb{S}^{d}$, a.e. $(x, t) \in Q$.
\smallskip
\item[(b)]
$c(x,t) = (c_{ijkl}(x,t))$ with $c_{ijkl} = c_{jikl} = c_{lkij} \in L^2(0, T; L^{\infty}(\Omega))$.
}

\medskip

\noindent
$\underline{H({\mathcal C}_e)}:$ \quad 
$\displaystyle {\mathcal C}_e \colon Q \times \real \to \mathbb{S}^{d}$ is such that 
\smallskip

\lista{
\item[(a)]
${\mathcal C}_e(\cdot,\cdot, r)$ is measurable on $Q$ for all $r \in \real$.
\smallskip
\item[(b)]
$\displaystyle 
\| {\mathcal C}_e(x, t, r) \|_{\mathbb{S}^{d}} \le c_{0e}(x,t) + c_{1e} \, | r |$ 
for all $r \in \real$, a.e. $(x, t) \in Q$ with $c_{e0} \in L^2(Q)$, $c_{e0}$, $c_{e1} \ge 0$.
\smallskip
\item[(c)]
$\displaystyle 
\| {\mathcal C}_e (x, t, r_1) - {\mathcal C}_e (x, t, r_2) \|_{\mathbb{S}^{d}} \le L_e \, | r_1 - r_2 |$ 
for all $r_1$, $r_2 \in \real$, a.e. $(x, t) \in Q$ with $L_e > 0$. 
}

\medskip

The contact and frictional potentials $j_\nu$ and $j_\tau$ and the potential $j$ satisfy
the following hypotheses.

\medskip

\noindent
$\underline{H(j_\nu)}:$ \quad 
$j_\nu \colon \Sigma_C \times \real \to \real$ is such that

\lista{
\item[(a)]
$j_\nu(\cdot, \cdot, r)$ is measurable on $\Sigma_C$ for all $r \in \real$ and there exists 
$e_0 \in L^2(\Gamma_C)$ such that $j_\nu(\cdot, \cdot, e_0(\cdot)) \in L^1(\Sigma_C)$.
\smallskip
\item[(b)]
$j_\nu (\bx, t, \cdot)$ is locally Lipschitz on $\real$ for a.e. $(\bx, t) \in \Sigma_C$.
\smallskip
\item[(c)]
$| \partial j_\nu (\bx, t, r) | \le c_{0\nu} (\bx, t)+ c_{1\nu} | r |$ for all $r \in \real$,  
a.e. $(\bx, t) \in \Sigma_C$ with $c_{0\nu} \in L^\infty(\Sigma_C)$, 
$c_{0\nu}$, $c_{1\nu} \ge 0$.
\smallskip
\item[(d)]
$(\zeta_1 - \zeta_2) (r_1 - r_2) \ge - m_\nu | r_1 - r_2 |^2$ for all 
$\zeta_i \in \partial j_\nu (\bx, t, r_i)$, $r_i \in \real$, $i =1$, $2$, 
a.e. $(\bx, t) \in \Sigma_C$ with $m_\nu \ge 0$.
}

\medskip

\noindent
$\underline{H(j_\tau)}:$ \quad 
$j_\tau \colon \Sigma_C \times \mathbb{R}^d \to \mathbb{R}$ is such that

%\smallskip

\lista{
\item[(a)]
$j_\tau (\cdot, \cdot, \bxi)$ is measurable on $\Sigma_C$ for all 
$\bxi \in \mathbb{R}^d$ and there exists ${\bf e}_1 \in L^2(\Gamma_C; \real^d)$ 
such that $j_\tau (\cdot, \cdot, {\bf e}_1(\cdot)) \in L^1(\Sigma_C)$. 
\smallskip
\item[(b)]
$j_\tau (\bx, t,\cdot)$ is locally Lipschitz on $\real^d$ for a.e. $(\bx, t) \in \Sigma_C$.
\smallskip
\item[(c)]
$\| \partial j_\tau(\bx, t, \bxi) \|_{\mathbb{R}^d} \le c_{0\tau} (\bx, t) 
+ c_{1\tau} \| \bxi \|_{\mathbb{R}^d}$ for all $\bxi \in \mathbb{R}^d$, 
a.e. $(\bx, t) \in \Sigma_C$ with $c_{0\tau} \in L^\infty(\Sigma_C)$, 
$c_{0\tau}$, $c_{1\tau} \ge 0$.
\smallskip
\item[(d)]
$(\bzeta_1 - \bzeta_2) \cdot (\bxi_1 - \bxi_2) \ge -
m_\tau \| \bxi_1 - \bxi_2 \|^2_{\mathbb{R}^d}$ for all 
$\bzeta_i \in \partial j_\tau (\bx, t, \bxi_i)$, $\bxi_i \in \mathbb{R}^d$, $i =1$, $2$, 
a.e. $(\bx, t) \in \Sigma_C$ with $m_\tau \ge 0$.
}

\medskip

\noindent
$\underline{H(j)}:$ \quad 
$j \colon \Sigma_C \times \mathbb{R} \to \mathbb{R}$ is such that

%\smallskip

\lista{
\item[(a)]
$j(\cdot, \cdot, r)$ is measurable on $\Sigma_C$ for all $r \in \mathbb{R}$ 
and there exists $e_2 \in L^2(\Gamma_C)$ such that 
$j(\cdot, \cdot, e_2(\cdot)) \in L^1(\Sigma_C)$.
\smallskip
\item[(b)]
$j (\bx, t, \cdot)$ is locally Lipschitz on $\real$ for a.e. $(\bx, t) \in \Sigma_C$.
\smallskip
\item[(c)]
$| \partial j (\bx, t, r) | \le c_0 (x, t)+ c_1 | r |$ for all $r \in \real$, 
a.e. $(\bx, t) \in \Sigma_C$ with $c_0 \in L^\infty(\Sigma_C)$, $c_0$, $c_1 \ge 0$.
\smallskip
\item[(d)] 
$(\zeta_1 - \zeta_2) (r_1 - r_2) \ge - m_0 | r_1 - r_2 |^2$ for all 
$\zeta_i \in \partial j (\bx, t, r_i)$, $r_i \in \real$, $i =1$, $2$, 
a.e. $(\bx, t) \in \Sigma_C$ with $m_0 \ge 0$.
}

\medskip

The thermal conductivity operator $K$, the operator $R$ in the heat equation, 
and the tangential function $h_\tau$ satisfy the following assumptions.

\medskip

\noindent
$\underline{H(K)}:$ \quad 
$\displaystyle K \colon Q \times \real^d \to \real^d$ is such that 

\lista{
\item[(a)]
$K(\cdot,\cdot, \xi)$ is measurable on $Q$ for all $\xi \in \real^d$.
\smallskip
\item[(b)]
$K(x, t, \cdot)$ is continuous on $\real^d$ for a.e. $(x, t) \in Q$.
\smallskip
\item[(c)]
$\displaystyle 
\| K(x, t, \xi) \|_{\real^d} \le k_0(x,t) + k_1 \, \| \xi \|_{\real^d}$ for all 
$\xi \in \real^d$, a.e. $(x, t) \in Q$ with $k_0 \in L^2(Q)$, $k_0 \ge 0$, 
$k_1 > 0$.
\smallskip
\item[(d)]
$\displaystyle 
\left( K(x, t, \xi_1) - K (x, t, \xi_2) \right) \cdot (\xi_1 - \xi_2) \ge 
m_K \, \| \xi_1 - \xi_2 \|^2_{\mathbb{R}^d}$
for all $\xi_1$, $\xi_2 \in \real^d$, a.e. $(x, t) \in Q$ with $m_K > 0$.  
\smallskip
\item[(e)]
$\displaystyle K(x, t, \xi) \cdot \xi \ge \alpha_K \, \| \xi \|^2_{\real^d}$ 
for all $\xi \in \real^d$, a.e. $(x, t) \in Q$ with $\alpha_K > 0$. 
}

\medskip

\noindent
$\underline{H(R)}:$ \quad 
$\displaystyle R \colon Q \times E \to L^2(\Omega)$ is such that 

\lista{
\item[(a)]
$R(\cdot, \cdot, v) \in L^2(Q)$ for all $v \in E$.
\smallskip
\item[(b)]
$\displaystyle 
\| R( x, t, v_1) - R (x, t, v_2) \|_{L^2(\Omega)} \le L_R \, \| v_1 - v_2 \|_E$ 
for all $v_1$, $v_2 \in E$, a.e. $(x, t) \in Q$ with $L_R > 0$. 
}

\medskip

\noindent
$\underline{H(h_\tau)}:$ \quad 
$\displaystyle h_\tau \colon \Gamma_C \times \real_+ \to \real_+$ is such that 

\lista{
\item[(a)]
$h_\tau(\cdot, r) \in L^2(\Gamma_C)$ for all $r \in \real_+$;
\smallskip
\item[(b)]
$\displaystyle 
| h_\tau (x, r_1) - h_\tau (x, r_2) | \le L_\tau \, | r_1 - r_2 |$
for all $r_1$, $r_2 \in \real_+$, a.e. $x \in \Gamma_C$ with $L_\tau > 0$. 
}

\medskip

We assume that the body forces, surface tractions, the density of heat sources 
and the initial conditions have the following regularity. 

\medskip

\noindent $\underline{H(f)}:$ \quad 
$\displaystyle 
f_0 \in L^2(0, T; E^*)$, $f_1 \in L^2(0, T; L^2(\Gamma_N; \real^d))$, 
$g \in L^2(0, T; V^*)$, $u_0 \in E$, 

\qquad\quad 
$v_0 \in H$ and $\theta_0 \in L^2(\Omega)$.

\section{Variational formulation of the problem}\label{Weak}
%%%%%%%%%%%%%%%%%%%%%%%%%%%%%

\noindent 
In this section, we obtain the variational formulation of Problem~$P$, 
establish the properties of the operators involved in the problem and 
formulate the main result on the unique solvability of Problem~$P$. 

First, we define the function $f \colon (0, T) \to E^*$ 
by 
\begin{equation}\label{force}
\langle f(t), v \rangle_{E^* \times E} = \langle f_0(t), v \rangle_{E^* \times E} + 
\langle f_1(t), v \rangle_{L^2(\Gamma_N; \real^d)} \ \ \mbox{for} \ \ v \in E 
\ \mbox{and a.e.} \ t \in (0, T). 
\end{equation}

\noindent 
Note that under the hypothesis $H(f)$, we have $f \in {\mathcal E}^*$. 
Assume that $(u, \sigma, \theta)$ is a triple of sufficiently smooth 
functions which solve Problem~$P$, $v \in E$ and $t \in (0, T)$.
We multiply the equation of motion (\ref{EQ1}) by $v$ and 
use the Green formula (\ref{green2}) to find that
\begin{equation}\label{NEWNR} 
\langle u''(t), v \rangle_{E^* \times E} + 
\langle \sigma(t), \varepsilon (v) \rangle_{\mathcal H} = 
\langle f_0(t), v \rangle_{E^* \times E} + \int_{\Gamma} \sigma(t) \nu \cdot v \, d\Gamma. 
\end{equation}

\noindent
We take into account the boundary conditions (\ref{EQ5}) 
and the fact that $v = 0$ on $\Gamma_D$ to obtain
\begin{equation}\label{bdry1}
\int_{\Gamma} \bsigma(t)\bnu\cdot \bv \, d\Gamma = 
\int_{\Gamma_N} f_1(t) \cdot \bv \, d \Gamma + 
\int_{\Gamma_C} \left( \sigma_\nu(t)v_\nu + 
\bsigma_\tau(t) \cdot \bv_\tau \right) \, d\Gamma.
\end{equation}

\noindent On the other hand, from the definition of the Clarke
subdifferential combined with (\ref{EQ6}), 
we have
\begin{equation*}
- \sigma_\nu(t) v_\nu \le j_\nu^0 (t, u_\nu'(t); v_\nu), \quad
- \bsigma_\tau(t) \cdot \bv_\tau \le j_\tau^0 (t, \bu_\tau'(t); \bv_\tau) 
\ \ \mbox{on} \ \ \Sigma_C, 
\end{equation*}

\noindent 
which implies 
\begin{equation}\label{bdry2}
\int_{\Gamma_C} \left( \sigma_\nu(t) v_\nu +
\bsigma_\tau(t) \cdot \bv_\tau \right) \, d\Gamma 
\ge -\int_{\Gamma_C} \Big( j_\nu^0 (t, u_\nu'(t); v_\nu)
+ j_\tau^0 (t, \bu_\tau'(t); \bv_\tau) \Big) \, d\Gamma.
\end{equation}

\noindent 
We now combine (\ref{force})--(\ref{bdry2}) to see
that
\begin{eqnarray}\label{first1}
&&
\hspace{-1.0cm}
\langle \bu''(t), \bv \rangle_{E^* \times E} +
\langle \bsigma (t), \bvarepsilon (\bv) \rangle_{\mathcal H} + 
\int_{\Gamma_C} \Big( j_\nu^0 (t, u_\nu'(t); v_\nu) + 
j_\tau^0 (t, \bu_\tau'(t); \bv_\tau) \Big)
\, d\Gamma \ge \\[2mm]
&&
\qquad \ge \langle \fb(t), \bv \rangle_{E^* \times E} \ \ \mbox{for all} \ \
v \in E \ \mbox{and a.e.} \ t \in (0, T).\nonumber
\end{eqnarray}

\noindent 
Next, we use (\ref{first1}) and the constitutive law (\ref{EQ2}) to obtain the
following inequality
\begin{eqnarray}
&& \label{inequality1}
\hspace{-1.0cm}
\langle u''(t) + A(t, u'(t)) + B(t, u(t)) + \int_0^t C(t-s) u(s) \, ds 
+ C_1(t, \theta (t)), v \rangle_{E^* \times E} + \\[2mm]
&& \qquad + \int_{\Gamma_C} \Big( j_\nu^0 (x, t, u_\nu'(t); v_\nu) + 
j_\tau^0 (t, u_\tau'(t); v_\tau) \Big) \, d\Gamma \ge \langle \fb(t), \bv \rangle_{E^* \times E}
\nonumber
\end{eqnarray}

\noindent 
for all $v \in E$ and a.e. $t \in (0, T)$, where the operators
$A$, $B$, $C \colon (0, T) \times E \to E^*$ and 
$C_1 \colon (0, T) \times L^2(\Omega) \to E^*$ 
are defined by
\begin{eqnarray}
&& \label{operaA}
\langle A(t, u), v \rangle_{E^* \times E} = 
\langle {\mathcal A} (t, \varepsilon (u)), \varepsilon (v) \rangle_{\mathcal H} 
\ \ \mbox{for all} \ \ u, v \in E, \\[2mm]
&& \label{operaB}
\langle B(t, u), v \rangle_{E^* \times E} = 
\langle {\mathcal B} (t, \varepsilon (u)), \varepsilon (v) \rangle_{\mathcal H} 
\ \ \mbox{for all} \ \ u, v \in E, \\[2mm]
&& \label{operaC}
\langle C(t) u, v \rangle_{E^* \times E} = 
\langle {\mathcal C} (t) \varepsilon (u)), \varepsilon (v) \rangle_{\mathcal H} 
\ \ \mbox{for all} \ \ u, v \in E, \\[2mm]
&& \label{operaC1}
\langle C_1(t, \theta), v \rangle_{E^* \times E} = 
\langle {\mathcal C}_e (t, \theta), \varepsilon (v) \rangle_{\mathcal H} 
\ \ \mbox{for all} \ \ u \in E, \, \theta \in L^2(\Omega) , 
\end{eqnarray}

\medskip

\noindent 
for a.e. $t \in (0, T)$. 
Next, let $\zeta \in V$ and $t \in (0, T)$. 
Multiplying the equation (\ref{EQ3}) by $\zeta$, 
using (\ref{EQ8}) and the Green formula (\ref{green1}), we have 
\begin{eqnarray}
\label{THETA1}
&&
\langle \theta'(t), \zeta \rangle_{V^* \times V} + 
\int_\Omega K(x, t, \nabla \theta(t)) \cdot \nabla \zeta \, dx 
- \int_{\Gamma_C} \frac{\partial \theta}{\partial \nu_K} \, \zeta \, d\Gamma = \\ [2mm] 
&& 
= \langle R(t, u'(t)) + g(t), \zeta \rangle_{V^* \times V}. \nonumber 
\end{eqnarray}

\noindent 
From the definition of the Clarke subdifferential and the condition (\ref{EQ7}), 
it follows that 
\begin{equation}\label{bdry3}
- \int_{\Gamma_C} \frac{\partial \theta}{\partial \nu_K} \, \zeta \, d\Gamma 
\le \int_{\Gamma_C} j^0(t, \theta (t); \zeta) \, d\Gamma - 
\int_{\Gamma_C} h_\tau (\| u_\tau'(t) \|_{\real^d}) \, \zeta \, d\Gamma. 
\end{equation}

\noindent 
By (\ref{THETA1}) and (\ref{bdry3}), we deduce the following inequality 
\begin{eqnarray}
&& \label{inequality2}
\hspace{-1.0cm}
\langle \theta'(t) + C_2(t, \theta(t)), \zeta \rangle_{V^* \times V} + 
\int_{\Gamma_C} j^0(t, \theta (t); \zeta) \, d\Gamma \ge 
\langle C_3(t, u' (t)) + g(t), \zeta \rangle_{V^* \times V} 
\end{eqnarray}

\noindent 
for all $\zeta \in V$ and a.e. $t \in (0, T)$, where the operators
$C_2 \colon (0, T) \times V \to V^*$ and $C_3 \colon (0, T) \times E \to V^*$ 
are given by
\begin{eqnarray}
&& \label{operaC2}
\langle C_2(t, \theta), \zeta \rangle_{V^* \times V} = 
\langle K (x, t, \nabla \theta), \nabla \zeta \rangle_{L^2(\Omega)} 
\ \ \mbox{for all} \ \ \theta, \, \zeta \in V, \\[2mm]
&& \label{operaC3}
\langle C_3 (t, v), \zeta \rangle_{V^* \times V} = 
\langle R(t, v), \zeta \rangle_{V^* \times V} + \int_{\Gamma_C}
h_\tau (\| v_\tau \|_{\real^d}) \, \zeta \, d\Gamma 
\end{eqnarray}

\noindent 
for all $v \in E$, $\zeta \in V$ and a.e. $t \in (0, T)$. 
Finally, we use (\ref{inequality1}), (\ref{inequality2}) and 
the initial conditions (\ref{EQ9}) to obtain the following 
system of hemivariational inequalities which is the 
variational formulation of Problem $P$.

\medskip

\noindent 
{\bf Problem} $P_V$:
find $u \in {\mathcal E}$ with $u' \in {\mathbb{E}}$ and $\theta \in {\mathcal W}$ 
such that 
\begin{eqnarray*}
&& 
\langle u''(t) + A(t, u'(t)) + B(t, u(t)) + \int_0^t C(t-s) u(s) \, ds 
+ C_1(t, \theta (t)), v \rangle_{E^* \times E} + \\[2mm]
&& 
\qquad + \int_{\Gamma_C} \Big( j_\nu^0 (x, t, u_\nu'(t); v_\nu) + 
j_\tau^0 (t, u_\tau'(t); v_\tau) \Big) \, d\Gamma \ge \langle f(t), v \rangle_{E^* \times E} 
\\ [2mm] 
&& 
\qquad \mbox{for all} \ v \in E \ \mbox{and a.e.} \ t \in (0, T) \\[2mm]
&&
\langle \theta'(t) + C_2(t, \theta(t)), \zeta \rangle_{V^* \times V} + 
\int_{\Gamma_C} j^0(t, \theta (t); \zeta) \, d\Gamma \ge 
\langle C_3(t, u' (t)) + g(t), \zeta \rangle_{V^* \times V} \\ [2mm] 
&& 
\qquad \mbox{for all} \ \zeta \in V \ \mbox{and a.e.} \ t \in (0, T) \\[2mm]
&& 
u(0) = u_0, \ \ u'(0) = v_0, \ \ \theta(0) = \theta_0 .
\end{eqnarray*}

In what follows we establish the properties of the operators 
involved in Problem~$P_V$. For the proofs of Lemmata~\ref{LEMMAA}, 
\ref{LEMMAB} and~\ref{LEMMAC}, we refer to Lemmata~8, 9 and~10, 
respectively, in~\cite{Kulig2}.

\begin{Lemma}\label{LEMMAA}
Under the hypothesis $H({\mathcal A})$, the operator $A \colon (0, T) \times E \to E^*$ 
defined by (\ref{operaA}) satisfies the properties

%%\smallskip

\lista{
\item[(a)]
$\displaystyle A(\cdot, v)$ is measurable on $(0, T)$ for all $v \in E$. 
\smallskip
\item[(b)]
$\displaystyle A(t, \cdot)$ is strongly monotone for a.e. $t \in (0,T)$, i.e. 
$\langle A (t, v) - A(t, u), v - u \rangle_{E^* \times E} \ge 
m_{\mathcal A} \| v - u \|_E^2$ for all $u$, $v \in E$,
a.e. $t \in (0, T)$. 
\smallskip
\item[(c)]
$\| A(t, v) \|_{E^*} \le {\widetilde{a}}_0(t) + {\widetilde{a}}_1 \| v \|_E$ 
for all $v \in V$, a.e. $t \in (0,T)$ with ${\widetilde{a}}_0 \in L^2(0,T)$, 
${\widetilde{a}}_0 \ge 0$ and ${\widetilde{a}}_1 >0$. 
\smallskip
\item[(d)]
$\langle A(t, v), v \rangle_{E^* \times E} \ge \alpha_{\mathcal A} \| v \|_E^2$ 
for all $v \in E$, a.e. $t \in (0,T)$. 
\smallskip
\item[(e)]
$\displaystyle A(t, \cdot)$ is pseudomonotone for a.e. $t \in (0,T)$,
}

\noindent 
where  
${\widetilde{a}}_0(t) = \sqrt{2} \, \| a_0 (t) \|_{L^2(\Omega)}$ and 
${\widetilde{a}}_1 = \sqrt{2} \, a_1$.
\end{Lemma}

\begin{Lemma}\label{LEMMAB}
Under the hypothesis $H({\mathcal B})$, 
the operator $B \colon (0,T) \times E \to E^*$ defined 
by (\ref{operaB}) satisfies the properties 

\lista{
\item[(a)]
$\displaystyle B(\cdot, v)$ is measurable on $(0, T)$ for all $v \in E$.  
\smallskip
\item[(b)]
$\displaystyle B(t, \cdot)$ is Lipschitz continuous for a.e. $t \in (0, T)$, i.e. 
$\| B(t, u) - B(t, v) \|_{E^*} \le L_B \| u - v \|_E$ for all $u$, $v \in E$, 
a.e. $t \in (0, T)$. 
\smallskip
\item[(c)]
$\| B(t, v) \|_{E^*} \le {\widetilde{b}}_0(t) + {\widetilde{b}}_1 \| v \|_E$ for all $v \in E$, 
a.e. $t \in (0,T)$ with ${\widetilde{b}}_0 \in L^2(0,T)$ and ${\widetilde{b}}_0$, 
${\widetilde{b}}_1 \ge 0$.
}

\noindent 
where  
${\widetilde{b}}_0(t) = \sqrt{2} \, \| b_0 (t) \|_{L^2(\Omega)}$ and 
${\widetilde{b}}_1 = \sqrt{2} \, b_1$.
\end{Lemma}

\begin{Lemma}\label{LEMMAC}
Under the hypothesis $H({\mathcal C})$, the operator $C$ defined 
by (\ref{operaC}) satisfies $C\in L^2(0,T; {\mathcal L}(E, E^*))$. 
\end{Lemma}

The proofs of Lemmata~\ref{LEMMAC1} and~\ref{LEMMAC3} are elementary 
and therefore they are omitted.

\begin{Lemma}\label{LEMMAC1}
Under the hypothesis $H({\mathcal C}_e)$, the operator 
$C_1 \colon (0,T) \times L^2(\Omega) \to E^*$ defined 
by (\ref{operaC1}) satisfies the properties 

\smallskip

\lista{
\item[(a)]
$\displaystyle C_1(\cdot, \theta)$ is measurable on $(0, T)$ for all $\theta \in L^2(\Omega)$. 
\smallskip
\item[(b)]
$\displaystyle C_1(t, \cdot)$ is Lipschitz continuous for a.e. $t \in (0, T)$, i.e. 
$\| C_1(t, \theta_1) - C_1(t, \theta_2) \|_{E^*} \le L_e \| \theta_1 - \theta_2 \|_{L^2(\Omega)}$ 
for all $\theta_1$, $\theta_2 \in L^2(\Omega)$, 
a.e. $t \in (0, T)$.  
\smallskip
\item[(c)]
$\| C_1(t, \theta) \|_{E^*} \le {\widetilde{c}}_{0e}(t) 
+ {\widetilde{c}}_{1e} \| \theta \|_{L^2(\Omega)}$ for all $\theta \in L^2(\Omega)$, 
a.e. $t \in (0,T)$ with ${\widetilde{c}}_{0e} \in L^\infty(0,T)$ and 
${\widetilde{c}}_{0e}$, ${\widetilde{c}}_{1e} \ge 0$.
}

\noindent 
where  
${\widetilde{c}}_{0e}(t) = \sqrt{2} \, \| c_{0e} (t) \|_{L^\infty(\Omega)}$ 
and 
${\widetilde{c}}_{1e} = \sqrt{2} \, c_{1e}$.
\end{Lemma}

\begin{Lemma}\label{LEMMAC2}
Under the hypothesis $H(K)$, the operator $C_2 \colon (0, T) \times V \to V^*$ 
defined by (\ref{operaC2}) satisfies the properties

\smallskip

\lista{
\item[(a)]
$\displaystyle C_2(\cdot, \theta)$ is measurable on $(0, T)$ for all $\theta \in V$. 
\smallskip
\item[(b)]
$\displaystyle C_2(t, \cdot)$ is strongly monotone for a.e. $t \in (0,T)$, i.e. 
there exists $m_1 > 0$ such that 
$\langle C_2 (t, \theta_1) - C_2(t, \theta_2), \theta_1 - \theta_2 \rangle_{V^* \times V} 
\ge m_K \| \theta_1 - \theta_2 \|_V^2$ for all $\theta_1$, $\theta_2 \in V$. 
\smallskip
\item[(c)]
$\| C_2(t, \theta) \|_{V^*} \le {\widetilde{k}}_0(t) 
+ {\widetilde{k}}_1 \| \theta \|_V$ for all $\theta \in V$, a.e. $t \in (0,T)$ 
with ${\widetilde{k}}_0 \in L^2(0,T)$, ${\widetilde{k}}_0 \ge 0$ 
and ${\widetilde{k}}_1 > 0$. 
\smallskip
\item[(d)]
$\langle C_2(t, \theta), \theta \rangle_{V^* \times V} \ge \alpha_K \, \| \theta \|_V^2$ 
for all $\theta \in V$, a.e. $t \in (0,T)$. 
\smallskip 
\item[(e)]
$\displaystyle C_2(t, \cdot)$ is pseudomonotone for a.e. $t \in (0,T)$,
}

\noindent 
where  
${\widetilde{k}}_0(t) = \sqrt{2} \, \| k_0 (t) \|_{L^\infty(\Omega)}$ 
and 
${\widetilde{k}}_1 = \sqrt{2} \, k_1$.
\end{Lemma}

\noindent 
{\bf Proof.} \ 
The properties (a)--(d) are direct consequences of the hypothesis $H(K)$. 
For the proof of (e), we apply Proposition 26.12 of~\cite[p.572]{ZEIDLER}
to deduce that the operator $C_2(t, \cdot)$ is monotone, coercive, bounded 
and continuous. In particular, it is monotone and hemicontinuous, so 
by Proposition~27.7(a) of~\cite[p.586]{ZEIDLER}, we infer that 
$C_2(t, \cdot)$ is pseudomonotone for a.e. $t \in (0,T)$.
{\hfill $\square$}

\begin{Lemma}\label{LEMMAC3}
Under the hypotheses $H(R)$ and $H(h_\tau)$, the operator 
$C_3 \colon (0,T) \times E \to V^*$ defined 
by (\ref{operaC3}) satisfies the properties 

\lista{
\item[(a)]
$\displaystyle C_3(\cdot, v)$ is measurable on $(0, T)$ for all $v \in E$.  
\smallskip
\item[(b)]
$\displaystyle C_3(t, \cdot)$ is Lipschitz continuous for a.e. $t \in (0, T)$, i.e. 
$\| C_3(t, v_1) - C_3(t, v_2) \|_{V^*} \le L_R \| v_1 - v_2 \|_E$ 
for all $v_1$, $v_2 \in E$, a.e. $t \in (0, T)$.
\smallskip
\item[(c)]
$\| C_3 (t, v) \|_{V^*} \le c_{31}(t) + c_{32} \, \| v \|_E$ for all $v \in E$, 
a.e. $t \in (0,T)$ with $c_{31} \in L^2(0,T)$ and $c_{31}$, $c_{32} \ge 0$.
}
\end{Lemma}

%\noindent {\bf Proof.} \
%\noindent 
%The proof of the lemma is complete.
%$\hfill\square$
%\medskip

We state the properties of the potential 
$J \colon (0, T) \times L^2(\Gamma_C) \to \real$ 
defined by
\begin{equation}\label{J}
J(t, \theta) = \int_{\Gamma_C} j(x, t, \theta(x)) \, d\Gamma 
\ \ \mbox{for all} \ \ \theta \in L^2(\Gamma_C), \ \mbox{a.e.} \ \ t \in (0, T).
\end{equation}

The proof of the Lemma~\ref{PROPJ} below follows the lines 
of the proof of Lemma~3.1 of~\cite{MOS1} and Lemma~5 of~\cite{MIGORSKI7}.

\begin{Lemma}\label{PROPJ}
Under the hypothesis $H(j)$ the functional $J$ given by (\ref{J}) 
has the following properties:

\smallskip

\lista{ 
\item[(a)] 
$J(\cdot, \theta)$ is measurable on $(0, T)$ for all $\theta \in L^2(\Gamma_C)$ 
and $J(\cdot, 0) \in L^1(0, T)$.
\smallskip
\item[(b)] 
$J(t,\cdot)$ is locally Lipschitz on $L^2(\Gamma_C)$ 
(in fact, Lipschitz on bounded subsets of $L^2(\Gamma_C)$) for a.e. $t \in (0,T)$.
\smallskip
\item[(c)] 
$\| \partial J(t, \theta) \|_{L^2(\Gamma_C)} \le \| c_0(t) \|_{L^2(\Gamma_C)} 
+ c_1 \| \theta \|_{L^2(\Gamma_C)}$ for $\theta \in L^2(\Gamma_C)$, 
a.e. $t \in (0, T)$.
\smallskip
\item[(d)] 
$\langle z_1 - z_2, \theta_1 - \theta_2 \rangle_{L^2(\Gamma_C)} \ge 
-m_0 \, \| \theta_1 - \theta_2 \|^2_{L^2(\Gamma_C)}$ for all 
$z_i \in \partial J(t, \theta_i)$, $\theta_i \in L^2(\Gamma_C)$, 
$i = 1$, $2$, a.e. $t \in (0, T)$.
\smallskip
\item[(e)] 
for all $\theta$, $\zeta \in L^2(\Gamma_C)$ and a.e. $t \in (0, T)$, we have
\begin{equation*}
J^0(t, \theta; \zeta) \le \int_{\Gamma_C} j^0(x, t, \theta(x); \zeta(x)) \, d\Gamma.
\end{equation*}
}

\end{Lemma}

\medskip

Our main existence and uniqueness result for Problem $P_V$ is formulated below. 
We denote by ${\bar{c}}_e$ the embedding constant of $E$ into $Z$ 
and by $c_e$ the embedding constant of $V$ into $Y$.

\begin{Theorem}\label{T1} 
Under the hypotheses $H({\mathcal A})$, $H({\mathcal B})$, $H({\mathcal C})$, 
$H({\mathcal C}_e)$, $H(j_\nu)$, $H(j_\tau)$, $H(j)$, $H(f)$, $H(K)$, $H(R)$, 
$H(h_\tau)$, and the following conditions
\begin{equation}\label{REGULAR1}
\left.
\begin{array}{l}
\mbox{either} \ \  j_\nu (\bx, t, \cdot) \ \mbox{and} \ j_\tau (\bx, t, \cdot) \ \mbox{are regular}  \\ [2mm]
\mbox{or} \ - j_\nu (\bx, t, \cdot) \ \mbox{and} \ - j_\tau (\bx, t, \cdot) \ \mbox{are regular}
\end{array}
\right\}
\end{equation}
\begin{equation}\label{REGULAR2}
\mbox{either} \ \  j (\bx, t, \cdot) \ \mbox{or} \ -j (\bx, t, \cdot) \ \mbox{is regular} 
\end{equation}
\begin{equation}\label{K2CONST}
\ m_{\mathcal A} \ge \max \,\{ m_\nu, m_\tau \} \, {{\bar{c}}_e}^2 \, \| \gamma \|^2 
\end{equation}
\begin{equation}\label{K2CONSTn}
\ \alpha_{\mathcal A} > 6 \, \max\,\{ c_{1\nu}, c_{1\tau} \} \, {{\bar{c}}_e}^2 \, \| \gamma \|^2 
\end{equation}
\begin{equation}\label{K2CONSTHEAT}
\ m_K \ge m_0 \,  \, {c_e}^2 \, \| \gamma_s \|^2 
\end{equation}
\begin{equation}\label{K2CONSTHEAT22}
\ \alpha_K > c_1 \, {c_e}^2 \, \| \gamma_s \|^2 ,
\end{equation}

\noindent 
Problem $P_V$ has a unique solution 
$\{ u, \theta \}$ such that 
$u \in {\mathcal E}$, $u' \in {\mathbb{E}}$ and $\theta \in {\mathcal W}$.
\end{Theorem}

\noindent 

%%%%%%%%%%%%%%%%%%%%%%%
\section{Proof of Theorem~\ref{T1}}\label{Main}
%%%%%%%%%%%%%%%%%%%%%%%

\noindent 
The proof of Theorem~\ref{T1} will be carried out in several steps. It is based on recent 
arguments of first and second order hemivariational inequalities and a fixed point argument. 
In the proof we consider two auxiliary intermediate problems.

{\bf Step 1.} \ Let $\eta \in {\mathcal E}^*$ be given. 
We consider the following second order hemivariational inequality. 

\medskip

\noindent 
{\bf Problem} $P_1^\eta$:
find $u_\eta \in {\mathcal E}$ such that $u_\eta' \in {\mathbb{E}}$ and  
such that 
\begin{eqnarray*}
&& 
\langle u_\eta''(t) + A(t, u_\eta'(t)), v \rangle_{E^* \times E} + \\[2mm]
&& 
\qquad + \int_{\Gamma_C} \Big( j_\nu^0 (x, t, u_{\eta\nu}'(t); v_\nu) + 
j_\tau^0 (t, u_{\eta\tau}'(t); v_\tau) \Big) \, d\Gamma \ge 
\langle f(t) - \eta(t), v \rangle_{E^* \times E} 
\\ [2mm] 
&& 
\qquad \mbox{for all} \ v \in E \ \mbox{and a.e.} \ t \in (0, T) \\[2mm]
&&
u_\eta(0) = u_0, \ \ u_\eta'(0) = v_0 .
\end{eqnarray*}

The unique solvability of Problem~$P_1^\eta$ is established by our next lemma.

\begin{Lemma}\label{HVIHyperb}
For $\eta \in {\mathcal E}^*$, Problem $P_1^\eta$ has a unique solution 
$u_\eta \in {\mathcal E}$ such that $u_\eta' \in {\mathbb{E}}$. 
Moreover, if $u_i$ denotes the solution to Problem $P_1^\eta$ 
corresponding to $\eta = \eta_i \in {\mathcal E}^*$, $i=1$, $2$, then 
there exists $c > 0$ such that 
\begin{equation}\label{EST1}
\| u_1(t) - u_2(t) \|_E^2 \le c \int_0^t \| \eta_1(s) - \eta_2(s) \|_{E^*}^2 \, ds 
\ \ \mbox{\rm for all} \ \ t \in [0, T].
\end{equation}
\end{Lemma}

\noindent 
{\bf Proof.} \ 
It follows from the hypotheses $H({\mathcal A})$, $H(j_\nu)$, 
$H(j_\tau)$, (\ref{REGULAR1}), (\ref{K2CONST}) and (\ref{K2CONSTn}) 
that we are able to apply Theorem~8.6 in \cite{MOSBOOK} from which 
we infer that Problem $P_1^\eta$ has a unique solution 
$u_\eta \in {\mathcal E}$ such that $u_\eta' \in {\mathbb{E}}$. 
Exploiting the method used for evolution hemivariational inequalities  
in Theorem~5.17 of~\cite{MOSBOOK} 
(cf. (5.86) and (5.88) in \cite{MOSBOOK}), 
we are able to show (\ref{EST1}) and the following estimate
for the first-order derivatives
\begin{equation}\label{EST1BOOK}
\| u_1'(t) - u_2'(t) \|_E^2 \le c \int_0^t \| \eta_1(s) - \eta_2(s) \|_{E^*}^2 \, ds 
\ \ \mbox{\rm for all} \ \ t \in [0, T]. 
\end{equation}

\noindent
For details we refer to Chapter~5 of~\cite{MOSBOOK}.
This completes the proof of the lemma.  
{\hfill $\square$}

\medskip 

{\bf Step 2.} \ 
We use the displacement field $u_\eta$ obtained in Lemma~\ref{HVIHyperb} 
and consider the following first order hemivariational inequality.

\medskip

\noindent 
{\bf Problem} $P_2^\eta$:
find $\theta_\eta \in {\mathcal W}$ such that   
such that 
\begin{eqnarray*}
&&
\langle \theta_\eta'(t) + C_2(t, \theta_\eta(t)), \zeta \rangle_{V^* \times V} + 
\int_{\Gamma_C} j^0(t, \theta_\eta(t); \zeta) \, d\Gamma \ge 
\langle C_3(t, u_\eta' (t)) + g(t), \zeta \rangle_{V^* \times V} \\ [2mm] 
&& 
\qquad \mbox{\rm for all} \ \zeta \in V \ \mbox{\rm and a.e.} \ t \in (0, T) \\[2mm]
&& 
\theta_\eta(0) = \theta_0 .
\end{eqnarray*}

The following result ensures the existence and uniqueness 
of a solution to Problem~$P_2^\eta$.

\begin{Lemma}\label{HVIParab}
For $\eta \in {\mathcal E}^*$, Problem $P_2^\eta$ has a unique solution 
$\theta_\eta \in {\mathcal W}$. 
Moreover, if $\theta_i$ denotes the solution to Problem $P_2^\eta$ 
corresponding to $\eta = \eta_i \in {\mathcal E}^*$, $i=1$, $2$, then 
there exists $c > 0$ such that 
\begin{equation}\label{EST2}
\| \theta_1(t) - \theta_2(t) \|_{L^2(\Omega)}^2 \le c \int_0^t \| \eta_1(s) - \eta_2(s) \|_{E^*}^2 \, ds 
\ \ \mbox{\rm for all} \ \ t \in [0, T].
\end{equation}
\end{Lemma}

\noindent 
{\bf Proof.} \ 
The proof of the lemma will be done in four steps. 
Consider the following evolution inclusion associated with Problem $P_2^\eta$.
\begin{equation}\label{theta1}
\begin{cases}
\ \ \displaystyle \mbox{find} \ \theta \in {\mathcal W} \ \mbox{such that} 
\\
\ \ \displaystyle 
\theta'(t) + C_2(t, \theta(t)) + \gamma_s^* \, \partial J(t, \gamma_s \theta(t)) 
\ni C_3(t, u'(t)) + g(t) \ \ \mbox{for a.e.} \ \ t \in (0,T) 
\\
\ \ \theta(0) = \theta_0 .
\end{cases}
\end{equation} 

Step $1^0$. \ 
Under the hypotheses $H(j)$ and (\ref{REGULAR2}), we prove that
$\theta \in {\mathcal W}$ is a solution to Problem $P_2^\eta$ 
if and only if $\theta$ solves (\ref{theta1}).

Let $\theta \in {\mathcal W}$ be a solution to (\ref{theta1}), i.e. 
there exists $\xi \in {\mathcal Y^*}$ such that $\xi(t) = \gamma_s^* z(t)$, 
$z(t) \in \partial J(t, \gamma_s \theta(t))$ for a.e. $t \in (0, T)$  
and
\begin{equation}\label{theta2}
\theta'(t) + C_2(t, \theta(t)) + \xi(t) = C_3(t, u'(t)) + g(t) \ \ \mbox{for a.e.} \ \ t \in (0,T) .
\end{equation}

\noindent
By the definition of the subdifferential, we have
\begin{equation}\label{theta3}
\langle z(t), w \rangle_{L^2(\Gamma_C)} \le J^0(t, \gamma_s \theta(t); w) 
\ \ \mbox{for all} \ w \in L^2(\Gamma_C), \ \mbox{a.e.} \ t \in (0,T) .
\end{equation}

\noindent
Combining Lemma~\ref{PROPJ}(e), (\ref{theta2}) and (\ref{theta3}), we obtain
\begin{eqnarray*}
&&
\langle C_3(t, u'(t)) + g(t) - \theta'(t) - C_2(t, \theta(t)), \zeta \rangle = \\ [2mm]
&& 
= \langle \xi (t), \zeta \rangle_{Y^* \times Y} 
= \langle z (t), \gamma_s \zeta \rangle_{L^2(\Gamma_C)}  
\le J^0 (t, \gamma_s \theta(t); \gamma_s \zeta) \le 
\int_{\Gamma_C} j^0 (x, t, \theta(t); \zeta) \, d\Gamma
\end{eqnarray*}

\noindent 
for all $\zeta \in V$, a.e. $t \in (0, T)$. 
Hence, $\theta$ is a solution to Problem $P_2^\eta$. 

Vice versa, let $\theta$ be a solution to Problem $P_2^\eta$. 
We note that the regularity hypothesis (\ref{REGULAR2}) implies 
that either $J(t, \cdot)$ or $- J(t, \cdot)$ is regular for a.e. $t \in (0, T)$, 
and the inequality in Lemma~\ref{PROPJ}(e) holds with equality, 
cf. Clarke~\cite{CLARKE}. 
Using this equality, we obtain
$$
\displaystyle \langle \theta'(t) + C_2(t, \theta(t)) - 
C_3(t, u'(t)) - g(t), \zeta \rangle_{V^* \times V} + J^0(t, \gamma_s \theta(t); \gamma_s \zeta)
\ge 0
$$

\noindent for all $\zeta \in V$ and a.e. $t \in (0, T)$. 
By Proposition 2.1(i) of \cite{MOS1}, we have
$$
\displaystyle \langle 
C_3(t, u'(t)) + g(t) - \theta'(t) - C_2(t, \theta(t)), \zeta \rangle_{V^* \times V} 
\le (J \circ \gamma_s)^0 (t, \theta(t); \zeta)
$$

\noindent for all $\zeta \in V$ and a.e. $t \in (0, T)$. 
Using the definition of the subdifferential and Proposition~2.1(ii) of~\cite{MOS1}, 
the previous inequality implies
that
$$
\displaystyle 
C_3(t, u'(t)) + g(t) - \theta'(t) - C_2(t, \theta(t)) \in 
\partial (J \circ \gamma_s) (t, \theta(t)) = \gamma_s^* \partial J (t, \gamma_s \theta(t))
$$

\noindent 
for a.e. $t \in (0, T)$. Thus $\theta$ is a solution to (\ref{theta1}). 
This completes the proof of Step $1^0$.

\medskip

Step $2^0$. \ 
Under the hypotheses $H({\mathcal C}_3)$, $H(j)$, $H(K)$, 
$H(R)$, $H(h_\tau)$ and (\ref{REGULAR2}), 
we prove that the evolution inclusion (\ref{theta1}) 
has a unique solution $\theta \in {\mathcal W}$. 

The proof of this step follows from the argument of Theorem~7 of~\cite{MIGORSKI8}. 
First, we suppose temporarily that the initial condition $\theta_0 \in V$. 
Let ${\widehat{C}}_2 \colon {\mathcal V} \to {\mathcal V^*}$ be the Nemitsky 
operator corresponding to $C_2$ and defined by 
$({\widehat{C}}_2 \theta)(t) = C_2(t, \theta(t) + \theta_0)$ 
for $\theta \in {\mathcal V}$ and a.e. $t \in (0, T)$.
Let ${\mathcal N} \colon {\mathcal V} \to 2^{\mathcal V^*}$ be the multivalued 
Nemitsky operator corresponding to 
$\gamma_s^* \circ \partial J(t, \gamma_s \, \cdot)$, i.e. 
$$
\displaystyle 
{\mathcal N} \theta = 
\{ \, w \in {\mathcal Y^*} \mid 
w(t) \in \gamma_s^* \partial J(t, \gamma_s (\theta(t) + \theta_0)) 
\ \ \mbox{a.e.} \ t \in (0, T) \, \} \ \ \mbox{for} \ \theta \in {\mathcal V}. 
$$

\noindent 
Under these notation, the problem (\ref{theta1}) can be written 
as the operator inclusion: 
\begin{equation}\label{theta4}
\begin{cases}
\displaystyle 
\ \theta' + {\widehat{C}}_2 \, \theta + {\mathcal N} \, \theta \ni {\widehat{C}}_3 (u') + g \\ 
\ \theta(0) = 0 , 
\end{cases}
\end{equation} 

\noindent 
where ${\widehat{C}}_3 \colon {\mathcal E} \to {\mathcal V}^*$ is given by 
$({\widehat{C}}_3 z) (t) = C_3(t, z(t))$ for $z \in {\mathcal E}$. 
Note that $\theta \in {\mathcal W}$ is a solution to problem (\ref{theta1}) 
if and only if $\theta - \theta_0 \in {\mathcal W}$ solves (\ref{theta4}).

\medskip 

Let $L \colon D(L) \subset {\mathcal V} \to {\mathcal V^*}$ be the operator 
defined by $L \theta = \theta'$ with 
$D(L) = \{ \theta \in {\mathcal W} \mid \theta (0) = 0 \}$. 
It is known (see e.g. \cite{ZEIDLER}) that $L$ is densely 
defined maximal monotone operator. 
Let ${\mathcal F} \colon {\mathcal V} \to 2^{\mathcal V^*}$ 
be the operator given by 
${\mathcal F} \theta = {\widehat{C}}_2 \, \theta + {\mathcal N} \, \theta$
for $\theta \in {\mathcal V}$. Now, the problem (\ref{theta4}) is equivalent 
to 
$$
\mbox{find} \ \theta \in D(L) \ \mbox{such that} \ 
L \theta + {\mathcal F} \theta \ni {\widehat{C}}_3 (u') + g. 
$$

\noindent 
In order to prove the existence of a solution to the problem (\ref{theta4}), 
we show that the operator ${\mathcal F}$ is bounded, coercive 
and $L$-pseudomonotone. The proof of boundedness and $L$-pseudomonotonicity 
is quite similar to that given in Theorem~7 of~\cite{MIGORSKI8}. 
We show the coercivity of ${\mathcal F}$. 
To this end, from the equality
\begin{eqnarray*}
\langle {\widehat{C}}_2 \theta, \theta \rangle_{{\mathcal V}^* \times {\mathcal V}}
= 
\int_0^T \langle C_2(t, \theta(t) + \theta_0), \theta(t) + \theta_0 \rangle_{V^* \times V} \, dt
-
\int_0^T \langle C_2(t, \theta(t) + \theta_0), \theta_0 \rangle_{V^* \times V} \, dt
\end{eqnarray*}

\noindent
for $\theta \in {\mathcal V}$, 
using (c) and (d) of Lemma~\ref{LEMMAC2}, and the H\"older inequality, 
we obtain
\begin{equation}\label{COERC2}
\langle {\widehat{C}}_2 \theta, \theta \rangle_{{\mathcal V}^* \times {\mathcal V}} 
\ge \alpha_K \, \| \theta + \theta_0 \|^2_{\mathcal V}  - c \, \| \theta \|_{\mathcal V} - c
\ge \alpha_K \, \| \theta \|^2_{\mathcal V}  - c \, \| \theta \|_{\mathcal V} - c
\end{equation}

\noindent
with a positive constant $c > 0$.
Next, let $\theta \in {\mathcal V}$, $w \in {\mathcal N} \theta$. So $w \in {\mathcal Y^*}$, 
$w(t) = \gamma_s^* \xi(t)$ and $\xi(t) \in \partial J(t, \gamma_s (\theta(t) + \theta_0))$ 
for a.e. $t \in (0, T)$.
Exploiting Lemma~\ref{PROPJ}(c), the continuity of the embedding $V \subset Y$ 
and of the trace operator $\gamma_s$, 
it follows that
\begin{eqnarray*}
&&
\hspace{-1.0cm}
\langle w, z \rangle_{{\mathcal V}^* \times {\mathcal V}}
= \int_0^T \langle w(t), z(t) \rangle_{V^* \times V} \, dt 
= \int_0^T \langle \xi(t), \gamma_s z(t) \rangle_{L^2(\Gamma_C)} \, dt \le \\
&&
\le c_e \, \| \gamma_s \| \int_0^T \| \xi (t) \|_{L^2(\Gamma_C)} \| z (t) \|_V \, dt \le \\ [2mm]
&&
\le
c_e \, \| \gamma_s \| \int_0^T 
\left(
\| c_0 (t) \|_{L^2(\Gamma_C)} + c_1 \, c_e \| \gamma_s \| \| \theta (t) + \theta_0 \|_V
\right) \| z (t) \|_V \, dt \le \\ [2mm]
&&
\le 
c_e \, \| \gamma_s \| \, \| c_0 \|_{L^2(\Sigma_C)} \| z \|_{\mathcal V} +
c_1 \, c_e^2 \| \gamma_s \|^2 \| \theta + \theta_0 \|_{\mathcal V} \| z \|_{\mathcal V}  
\end{eqnarray*}

\noindent
for all $z \in {\mathcal V}$. Hence, we infer
$$
\displaystyle
\| w \|_{\mathcal V^*} \le 
c_e \, \| \gamma_s \| \, \| c_0 \|_{L^2(\Sigma_C)} +
c_1 \, c_e^2 \| \gamma_s \|^2 (\| \theta \|_{\mathcal V} + T \| \theta_0 \|_V) 
$$

\noindent 
and 
\begin{equation*}
| \langle {\mathcal N} \theta, \theta \rangle_{{\mathcal V^*} \times {\mathcal V}}| 
= | \langle w, \theta \rangle_{{\mathcal V^*} \times {\mathcal V}}| \le 
\| w \|_{\mathcal V^*} \| \theta \|_{\mathcal V} \le 
c_1 \, c_e^2 \| \gamma_s \|^2 \| \theta \|^2_{\mathcal V} + c \| \theta \|_{\mathcal V} 
\end{equation*}

\noindent 
with a positive constant $c$. The latter and (\ref{COERC2}) implies
\begin{eqnarray*}
\langle {\mathcal F} \theta, \theta \rangle_{{\mathcal V}^* \times {\mathcal V}}
= \langle {\widehat{C}}_2 \theta, \theta \rangle_{{\mathcal V}^* \times {\mathcal V}} 
+ \langle {\mathcal N} \theta, \theta \rangle_{{\mathcal V}^* \times {\mathcal V}} \ge 
(\alpha_K - c_1 \, c^2_e \, \| \gamma_s \|^2) \| \theta (t) \|_{\mathcal V} 
- c \, \| \theta \|_{\mathcal V} - c
\end{eqnarray*}

\noindent
Finally, by the hypothesis (\ref{K2CONSTHEAT22}), 
we deduce that the operator ${\mathcal F}$ is coercive.

\medskip

Since the multivalued operator ${\mathcal F}$ is bounded, coercive and $L$-pseudomonotone, 
from Theorem~6.3.73 in~\cite{DMP2}, it follows that the problem (\ref{theta4}) 
has a solution $\theta \in D(L)$, so $\theta + \theta_0$ solves (\ref{theta1}) in the case 
$\theta_0 \in V$. 
Subsequently, exploiting the method used in Theorem~7 of~\cite{MIGORSKI8}, 
we are able to prove that the problem (\ref{theta1}) has a solution $\theta \in {\mathcal W}$ 
in the case $\theta_0 \in L^2(\Omega)$. 

\medskip 

Step $3^0$. \ 
We claim that the solution to Problem $P_2^\eta$ is unique. From Step~$1^0$, 
it is enough to prove that the problem (\ref{theta1}) has a unique solution.
Let $\theta_1$, $\theta_2 \in {\mathcal W}$ be solutions to (\ref{theta1}), 
i.e. 
\begin{equation}\label{eq82}
\theta_1'(t) + C_2(t, \theta_1(t)) + \xi_1(t) = C_3(t, u'(t)) + g(t)  \ \ \mbox{a.e.} \ t \in (0, T),
\end{equation}
\begin{equation}\label{eq83}
\theta_2'(t) + C_2(t, \theta_2(t)) + \xi_2(t) = C_3(t, u'(t)) + g(t)  \ \ \mbox{a.e.} \ t \in (0, T),
\end{equation}
\begin{equation}\label{eq84}
\xi_1(t) \in \gamma_s^* \partial J(t, \gamma_s \theta_1(t)),
\ \ 
\xi_2(t) \in \gamma_s^* \partial J(t, \gamma_s \theta_2(t)),
\ \ \mbox{a.e.} \ t \in (0, T),
\end{equation}
\begin{equation}\label{eq85}
\theta_1(0) = \theta_2(0) = \theta_0 .
\end{equation}

\noindent 
Subtracting (\ref{eq83}) from (\ref{eq82}), multiplying the result by $\theta_1(t) - \theta_2(t)$ 
and integrating by parts on $[0, t]$ with the initial conditions (\ref{eq85}), we obtain
\begin{eqnarray}\label{eq86}
&&
\hspace{-1.3cm}
\frac{1}{2}\, \| \theta_1(t) - \theta_2(t)\|^2_{L^2(\Omega)} + 
\int_0^t \langle C_2(s, \theta_1(s)) - C_2(s, \theta_2(s)), \theta_1(s) 
- \theta_2(s) \rangle_{V^* \times V} \, ds + \\ [2mm]
\nonumber
&& 
+ \int_0^t \langle \xi_1(s) - \xi_2(s), \theta_1(s) - \theta_2(s) \rangle_{V^* \times V} \, ds = 0
\ \ \mbox{for all} \ \ t \in [0, T].
\end{eqnarray}

\noindent From (\ref{eq84}), we have 
$\xi_i (t) = \gamma_s^* z_i(t)$ with $z_i(t) \in \partial J(t, \gamma_s \theta_i(t))$ 
for a.e. $t \in (0, T)$ and $i=1$, $2$. 
By Lemma~\ref{PROPJ}(d), we deduce
\begin{eqnarray}\label{eq87}
&&
\int_0^t \langle \xi_1(s) - \xi_2(s), \theta_1(s) - \theta_2(s) \rangle_{V^* \times V} \, ds = 
\\ [2mm]
&&
\nonumber
= \int_0^t \langle z_1(s) - z_2(s), \gamma_s \theta_1(s) - \gamma_s \theta_2(s) 
\rangle_{L^2(\Gamma_C)} \, ds \ge \\ [2mm]
&&
\nonumber
\ge -m_0 \int_0^t \| \gamma_s \theta_1(s) - \gamma_s \theta_2(s) \|^2_{L^2(\Gamma_C)} \, ds 
\ge -m_0 \, c_e^2  \, \| \gamma_s \|^2 \int_0^t \| \theta_1(s) - \theta_2(s)\|_V^2 \, ds
\end{eqnarray}

\noindent 
for all $t\in[0,T]$. Inserting the inequality (\ref{eq87}) into (\ref{eq86}), 
using Lemma~\ref{LEMMAC2}(b) and (\ref{K2CONSTHEAT}), we obtain
$$
\displaystyle 
\frac{1}{2}\, \| \theta_1(t) - \theta_2(t)\|^2_{L^2(\Omega)} + 
c \, \int_0^t \| \theta_1(s) - \theta_2(s) \|_V^2 \, ds \le 0
$$

\noindent 
for all $t \in [0,T]$ with $c = m_K - m_0 \, c_e^2 \, \| \gamma_s \|^2 \ge 0$. 
Hence we deduce that $\theta_1 = \theta_2$ which completes the proof 
of the uniqueness of solution.

\medskip 

Step $4^0$. \ 
We will establish the estimate (\ref{EST2}). Let $\eta_i \in {\mathcal E}^*$ 
and let $\theta_i = \theta_{\eta_i}$ be the unique solutions to Problem $P_2^\eta$ 
corresponding to $\eta_i$, $i=1$, $2$. We use the same technique as in Step $3^0$. 
Subtracting the equations satisfied by $\theta_i$, multiplying the result by 
$\theta_1(t) - \theta_2(t)$ and integrating on $[0, t]$, we deduce
\begin{eqnarray*}
&&
\hspace{-1.0cm}
\frac{1}{2}\, \| \theta_1(t) - \theta_2(t)\|^2_{L^2(\Omega)} + 
\int_0^t \langle C_2(s, \theta_1(s)) - C_2(s, \theta_2(s)), \theta_1(s) 
- \theta_2(s) \rangle_{V^* \times V} \, ds + \\ [2mm]
&& 
+ \int_0^t \langle \xi_1(s) - \xi_2(s), \theta_1(s) - \theta_2(s) \rangle_{V^* \times V} \, ds = \\ [2mm]
&&
= \int_0^t \langle C_3(s, u_1'(s)) - C_3(s, u_2'(s)), 
\theta_1(s) - \theta_2(s) \rangle_{V^* \times V} \, ds , 
\end{eqnarray*}

\noindent
where $\xi_i(t) = \gamma_s^* z_i(t)$, $z_i(t) \in \partial J(t, \gamma_s \theta_i(t))$
for a.e. $t \in (0, T)$, $i=1$, $2$. 
Exploiting Lemma~\ref{LEMMAC2}(b), Lemma~\ref{LEMMAC3}(b), (\ref{eq87}) 
and the Young inequality with $\varepsilon > 0$, we have
\begin{eqnarray*}
&&
\frac{1}{2}\, \| \theta_1(t) - \theta_2(t)\|^2_{L^2(\Omega)} + 
m_K \int_0^t \| \theta_1(s) - \theta_2(s) \|_V^2 \, ds \le \\ [2mm]
&& 
\le \int_0^t \| C_3(s, u_1'(s)) - C_3(s, u_2'(s)) \|_{V^*} 
\| \theta_1(s) - \theta_2(s) \|_V \, ds \le \\ [2mm]
&&
\le 
\frac{L_R}{2\varepsilon^2} \int_0^t \| u_1'(s) - u_2'(s) \|^2_E \, ds 
+ \frac{\varepsilon^2}{2} \int_0^t \| \theta_1(s) - \theta_2(s) \|^2_V \, ds
\end{eqnarray*}

\noindent 
for all $t \in [0,T]$. Choosing $\varepsilon = \sqrt{2 m_K}$, 
we conclude
\begin{equation*}
\frac{1}{2}\, \| \theta_1(t) - \theta_2(t)\|^2_{L^2(\Omega)} 
\le 
\frac{L_R}{4 m_K} \int_0^t \| u_1'(s) - u_2'(s) \|^2_E \, ds
\end{equation*}

\noindent 
for all $t \in [0,T]$. Finally, we use the estimate (\ref{EST1BOOK}) 
and the previous inequality to obtain (\ref{EST2}). 
This completes the proof of the lemma. 
{\hfill $\square$}

\medskip 

{\bf Step 3.} \ 
In this step, we apply a fixed point argument. Let $u_\eta \in {\mathcal E}$ 
with $u_\eta' \in {\mathbb{E}}$ be the solution to Problem $P_1^\eta$ and 
let $\theta_\eta \in {\mathcal W}$ be the solution to Problem $P_2^\eta$ 
obtained in Lemma~\ref{HVIHyperb} and Lemma~\ref{HVIParab}, respectively. 
We define the operator $\Lambda \colon {\mathcal E}^* \to {\mathcal E}^*$ 
by 
\begin{equation}\label{LLL1}
\langle \Lambda \eta (t), v \rangle_{E^* \times E} = 
\langle B(t, u_\eta(t)) + \int_0^t C(t-s) u_\eta(s) \, ds 
+ C_1(t, \theta_\eta (t)), v \rangle_{E^* \times E} 
\end{equation}

\noindent 
for all $v \in E$ and a.e. $t \in (0, T)$. 

\begin{Lemma}\label{Lambda}
The operator $\Lambda$ defined by (\ref{LLL1}) has 
a unique fixed point $\eta^* \in {\mathcal E}^*$.
\end{Lemma}

\noindent 
{\bf Proof.} \ 
It is easy to check that the operator $\Lambda$ is well defined. Indeed, 
from Lemmata~\ref{LEMMAB}$(c)$ and~\ref{LEMMAC1}$(c)$ 
and the inequality
\begin{equation*}
\|  \int_0^t C(t - s) u_\eta (s) \, ds \|_{E^*} \le 
\int_0^t \| C(t - s) \|_{{\mathcal L}(E, E^*)} \| u_\eta (s) \|_E \, ds \le
\end{equation*}
$$
\displaystyle 
\le \left( \int_0^t \| C(\tau) \|^2_{{\mathcal L}(E, E^*)} \, d\tau \right)^{1/2} 
\left( \int_0^t \| u_\eta(\tau) \|_E^2 \, d \tau \right)^{1/2} \le 
\| C \|_{L^2(0, t; {\mathcal L}(E, E^*))} \, \| u_\eta \|_{L^2(0, t; E)}
$$

\noindent for all $t \in [0, T]$, we have
\begin{eqnarray*}
&&
\hspace{-1.3cm}
\| \Lambda \eta \|^2_{\mathcal E^*} = 
\int_0^T \| (\Lambda \eta)(s) \|^2_{E^*} \, ds \le
c \int_0^T \Big( 
\| B(s, u_\eta(s)) \|_{E^*}^2 + \\ [2mm]
&&
+ \ \| \int_0^s C(t-s') u_\eta (s') \, ds' \|^2_{E^*} + 
\| C_1(s, \theta (s)) \|^2_{E^*} \Big) \, ds \le \\ [2mm]
&&
\le c \left(
1 + \| u_\eta \|^2_{\mathcal E} + \| \theta_\eta \|^2_{\mathcal V} \right) 
\end{eqnarray*}

\noindent 
where $c > 0$.
Hence 
$\| \Lambda \eta \|_{\mathcal E^*} \le 
c \left( 1 + \| u_\eta \|_{\mathcal E} + \| \theta_\eta \|_{\mathcal V} \right)$ 
which implies that the operator $\Lambda$ is well defined and takes 
values in ${\mathcal E^*}$.

Subsequently, we will show that the operator $\Lambda$ has a
unique fixed point. Let $\eta_1$, $\eta_2 \in {\mathcal E^*}$. 
By (\ref{LLL1}), we have 
\begin{eqnarray*}
&&
\| \Lambda \eta_1(t) - \Lambda \eta_2(t) \|^2_{E^*} 
\le c \Big(
\| B(t, u_1(t)) - B(t, u_2(t)) \|_{E^*}^2 + \\ [2mm]
&&
+ \ \| \int_0^t C(t-s) (u_1(s) - u_2(s)) \, ds \|^2_{E^*} + 
\| C_1(t, \theta_1 (t)) - C_1(t, \theta_2 (t)) \|^2_{E^*} \Big) .  
\end{eqnarray*}

\noindent 
Using Lemmata~\ref{LEMMAB}$(b)$ and~\ref{LEMMAC1}$(b)$, 
and the inequality
\begin{equation*}
\| \int_0^t C(t-s) (u_1(s) - u_2(s)) \, ds \|^2_{E^*} \le 
\| C \|^2_{L^2(0, T; {\mathcal L}(E, E^*))} \int_0^t \| u_1(s) - u_2 (s) \|^2_E \, ds
\end{equation*}

\noindent for all $t \in [0, T]$, we deduce
\begin{eqnarray*}
&&
\| \Lambda \eta_1(t) - \Lambda \eta_2(t) \|^2_{E^*} 
\le c \Big(
\| u_1(t) - u_2 (t) \|^2_E + \int_0^t \| u_1(s) - u_2 (s) \|^2_E \, ds + \\ [1mm]
&&
\hspace{3.8cm}
+ \| \theta_1 (t) - \theta_2 (t) \|^2_{L^2(\Omega)} \Big) .  
\end{eqnarray*}

\noindent
Hence, by (\ref{EST1}) and (\ref{EST2}), we obtain
\begin{equation*}
\| \Lambda \eta_1(t) - \Lambda \eta_2(t)\|^2_{E^*} \le 
c \int_0^t \| \eta_1(s) - \eta_2(s)\|^2_{E^*} \, ds 
\end{equation*}

\noindent 
for all $t \in [0, T]$ with $c > 0$. 
Applying Lemma~\ref{LemmaKulig}, we infer that 
there exists a unique $\eta^* \in {\mathcal E}^*$ 
such that $\Lambda \eta^* = \eta^*$. 
This completes the proof of the lemma. 
{\hfill $\square$}

\medskip

{\bf Step 4.} \ 
We have now all ingredients to prove the theorem.
%%\medskip
%%{\bf Proof of Theorem~\ref{T1}}. \ 
Let $\eta^* \in {\mathcal E}^*$ be the unique fixed point of the operator $\Lambda$ 
established in Lemma~\ref{Lambda}, i.e. 
\begin{equation*}
\eta^* (t) = B(t, u_{\eta^*} (t)) + 
\int_0^t C(t-s) u_{\eta^*}(s) \, ds + C_1(t, \theta_{\eta^*} (t)) 
\end{equation*}

\noindent 
for a.e. $t \in (0, T)$. Let $u^* = u_{\eta^*}$ be the unique solution 
of Problem~$P_1^{\eta^*}$ corresponding to $\eta^*$ established in Lemma~\ref{HVIHyperb}. 
Moreover, let $\theta^* = \theta_{\eta^*}$ be the unique solution 
of Problem~$P_2^{\eta^*}$ proved in Lemma~\ref{HVIParab}. 
Hence, $\{ u^*, \theta^* \}$ is the unique solution to Problem~$P_V$ 
with the regularity
$u^* \in {\mathcal E}$, $u^{*'} \in {\mathbb{E}}$ and $\theta^* \in {\mathcal W}$. 
The uniqueness part of the theorem is a consequence of the uniqueness 
of the fixed point of~$\Lambda$ and Lemmata~\ref{HVIHyperb} and~\ref{HVIParab}. 
This completes the proof of the theorem. 
{\hfill $\square$}

\section{Examples}\label{Examples}
%%%%%%%%%%%%%%%%%%

\noindent 
We now give a simple example of the functional which 
satisfies hypothesis $H(J)_1$. 

\begin{Example}\label{Example1}
Let us consider the functional 
$J_1 \colon L^2(\Gamma_C; \real^d) \to \real$ defined by 
$$ 
\displaystyle 
J_1(v) = \int_{\Gamma_C} \Big(  
\int_0^{v_N(x)} \beta(s) \, ds \Big) \, d\Gamma(x) \ \ \ 
\mbox{for all} \ v \in L^2(\Gamma_C; \real^d)
$$

\noindent 
(for simplicity we drop the $(x,t)$-dependence in the 
integrand of $J$), where the function $\beta$ satisfies 
the following hypothesis (cf. $H(p_N)$ in Section \ref{Statement}): 

\smallskip 
\smallskip

\noindent 
$\underline{H(\beta)}:$ \quad 
$\beta \in L^\infty_{loc}(\real)$ is a function such that 
$| \beta(s) | \le \beta_0 ( 1 + |s| )$ for $s \in \real$ 

\qquad \ \ 
with $\beta_0 > 0$, 
$\displaystyle 
\lim_{\tau\to \xi_{\pm}} \beta(\tau)$ exist for every 
$\xi \in \real$ and 
\begin{equation}\label{Hbeta}
\displaystyle 
\essinf_{\xi_1 \not= \xi_2} \, 
{{\beta(\xi_1)- \beta(\xi_2)} \over {\xi_1 - \xi_2}} 
\ge - m_2 \ \ \mbox{with some} \ m_2 > 0. 
\end{equation} 

\noindent 
We define the multivalued map 
$\displaystyle \widehat{\beta} \colon \real \to 2^{\real}$ 
which is obtained from $\beta$ 
by "filling in the gaps" at its discontinuity points, 
i.e. 
$\displaystyle \widehat{\beta}(\xi) = 
[ \underline{\beta}(\xi), \overline{\beta}(\xi) ]$, 
where
$$
\displaystyle 
\underline{\beta}(\xi) = 
\lim_{\delta\to 0^+} \essinf_{|t-\xi|\le\delta} \beta(t), 
\ \ \
\overline{\beta}(\xi) = 
\lim_{\delta\to 0^+} \esssup_{|t-\xi|\le\delta} \beta(t)
$$

\noindent
and $[\cdot,\cdot]$ denotes the interval. 
It is well known (see e.g. \cite{GMDR}) that 
a locally Lipschitz function $j_N \colon \real \to \real$ can be
determinated, up to an additive constant, by the relation
$j_N(s) = \int_0^{s} \beta(\tau) \, d\tau$ 
and 
$\displaystyle \partial j_N (s) = {\widehat{\beta}} (s)$ 
for $s \in \real$. 
It can be shown (see \cite{MIGORSKI7} for the details) that 
$j_N$ satisfies $H(j_N)$ and the functional $J_1$ satisfies $H(J)_1$. 
\end{Example}


\begin{thebibliography}{99} 

\baselineskip=12pt

\bibitem{ADLY} 
S. Adly, O. Chau and M. Rochdi, 
Solvability of a class of thermal dynamical contact problems with subdifferential 
conditions, {\it Numerical Algebra, Control and Optimization,} 2 (2012), 91--104.

\bibitem{AKRS} 
A. Amassad, K. L. Kuttler, M. Rochdi and M. Shillor, 
Quasi-static thermoviscoelastic contact problem with slip 
dependent friction coefficient, 
{\it Math. Comp. Modeling,} 36 (2002), 839--854.

\bibitem{A2002} 
K. T. Andrews, K. L. Kuttler, M. Rochdi and M. Shillor, 
One-dimensional dynamic thermoviscoelastic contact with damage,  
{\it J. Math. Anal. Appl.}, 272 (2002), 249--275.

\bibitem{A1997} 
K. T. Andrews, M. Shillor, S. Wright and A. Klarbring, 
A dynamic thermoviscoelastic contact problem with friction and wear,  
{\it Int. J. Engng Sci.,} 35 (1997), 1291--1309.

%\bibitem{AC}
%J.P. Aubin and A. Cellina,
%{\it Differential Inclusions. Set-Valued Maps and Viability Theo\-ry,}
%Springer-Verlag, Berlin, New York, Tokyo (1984).

%\bibitem{AUCL}
%J.-P. Aubin and F.H. Clarke, 
%Shadow prices and duality for a class of optimal 
%control problems, 
%{\it SIAM J. Control Optim.}, 17 (1979), 567--586.

%\bibitem{AWBI}
%B. Awbi, El H. Essoufi and M. Sofonea,
%A viscoelastic contact problem with normal 
%damped response and friction, 
%{\it Ann. Polon. Math.}, 75 (2000), 233--246. 

%\bibitem{BM}
%J. Berkovits and V. Mustonen,
%Monotonicity Methods for Nonlinear Evolution Equations,
%{\it Nonlinear Analysis,}
%27 (12) (1996), 1397--1405.

%\bibitem{BH}
%F.E. Browder and P. Hess,
%Nonlinear mappings of monotone type in Banach spaces, 
%{\it J. Funct. Anal.,} 11 (1972), 251--294.

%\bibitem{CHANG}
%K. C. Chang, 
%Variational methods for nondifferentiable functionals 
%and applications to partial differential equations,
%{\it J. Math. Anal. Appl.,} 80 (1981), 102--129.

%\bibitem{CHAU}
%O. Chau, W. Han and M. Sofonea,  
%A dynamic frictional contact problem with normal damped response, 
%{\it Acta Appl. Math.,} 71 (2002), 159--178. 

\bibitem{CHAU2} 
O. Chau, R. Oujja and M. Rochdi, 
A mathematical analysis of a dynamical frictional contact model 
in thermoviscoelasticity, 
{\it Discrete and Cont. Dyn. Systems, Ser. S,} 1 (2008), 61--70.

\bibitem{CLARKE} 
F. H. Clarke, 
{\it Optimization and Nonsmooth Analysis,} 
Wiley - Interscience, New York (1983).


\bibitem{DMTHERMO}
Z. Denkowski and S. Mig\'orski, 
A system of evolution hemivariational inequalities modeling 
thermoviscoelastic frictional contact, 
{\it Nonlinear Analysis,} 60 (2005), 1415--1441. 

\bibitem{Denkowski}
Z. Denkowski and S. Migorski, 
Hemivariational inequalities in thermoviscoelasticity, 
{\it Nonlinear Analysis,} 63 (2005), 87--97. 

\bibitem{DMO}
Z. Denkowski, S. Migorski and A. Ochal, 
Optimal control for a class of mechanical thermoviscoelastic 
frictional contact problems, 
{\it Control and Cybernetics,} 36 (2007), 611--632. 

\bibitem{DMP2} 
Z. Denkowski, S. Mig\'orski and N.S. Papageorgiou, 
{\it An Introduction to Nonlinear Analysis: Applications}, 
Kluwer/Plenum, New York (2003).  

%\bibitem{DUMONT}
%Y. Dumont, D. Goeleven, M. Rochdi, K.L. Kuttler and M. Shillor, 
%A dynamic model with friction and adhesion with applications to rocks, 
%{\it J. Math. Anal. Appl.}, 247 (2000), 87--109. 

\bibitem{DL}
G. Duvaut and J. L. Lions, 
{\it Inequalities in Mechanics and Physics},
Springer-Verlag, Berlin (1976). 

\bibitem{FT}
I. Figueiredo and L. Trabucho, 
A class of contact and friction dynamic problems in
thermoelasticity and in thermoviscoelasticity, 
{\it Int. J. Engng Sci.,} 33 (1995), 45--66.

%\bibitem{G}
%L. Gasi\'nski,   
%{\it Hyperbolic Hemivariational Inequalities and their Applications 
%to Optimal Shape Design,} 
%PhD Thesis, Jagiellonian University, Cracow, Poland (2000), 
%in Polish, p.61. 

%\bibitem{GS}
%L. Gasi\'nski and M. Smo{\l}ka,
%An existence theorem for wave-type hyperbolic hemivariational 
%inequalities, {\it Math. Nachr.}, 242 (2002), 1-12.   

\bibitem{GO11}
D. Goeleven, M. Miettinen and P.D. Panagiotopoulos, 
Dynamic hemivariational inequalities and their applications, 
{\it J. Optimiz. Theory and Appl.}, 103 (3) (1999), 567--601. 

%\bibitem{GMDR}
%D. Goeleven, D. Motreanu, Y. Dumont and M. Rochdi, 
%{\it Variational and Hemivariational Inequalities: Theory, 
%Methods and Applications}, 
%Kluwer Academic Publishers, Boston, Dordrecht, London (2003).  

%\bibitem{GOMO}
%D. Goeleven and D. Motreanu,  
%Hyperbolic hemivariational inequality and nonlinear wave equation 
%with discontinuities, in: {\it From Convexity to Nonconvexity}, 
%R.P. Gilbert et al. (Eds.), Kluwer (2001), 111--122.  

\bibitem{HAN}
W. Han and M. Sofonea, 
{\it Quasistatic Contact Problems in Viscoelasticity and Viscoplasticity}, 
American Mathematical Society, International Press (2002).

\bibitem{HMP}
J. Haslinger, M. Miettinen and P.D. Panagiotopoulos,  
{\it Finite Element Method for Hemivariational Inequalities. 
Theory, Methods and Applications,} 
Kluwer Academic Publishers, Boston, Dordrecht, London (1999). 

\bibitem{JAR}
J. Jarusek, 
Dynamic contact problems with given friction for viscoelastic bodies,  
{\it Czech. Math. J.}, 46 (1996), 475--487.

\bibitem{Kulig2}
A. Kulig, 
Hyperbolic hemivariational inequalities for dynamic viscoelastic 
contact problems, {\it Journal of Elasticity,} 110 (2013), 1--31.

\bibitem{Kulig1}
A. Kulig and S. Mig\'orski, 
Solvability and continuous dependence results for second order 
nonlinear evolution inclusions with a Volterra-type operator, 
{\it Nonlinear Analysis,} 75 (2012), 4729--4746. 

\bibitem{KUT}
K. L. Kuttler, 
Dynamic friction contact problem with general normal and friction laws, 
{\it Nonlinear Analysis}, 28 (1997), 559--575.

%\bibitem{KUTSHI}
%K. L. Kuttler and M. Shillor, 
%Set-valued pseudomonotone maps and degenerate evolution inclusions, 
%{\it Comm. Contemp. Math.}, 1 (1999), 87--123.

\bibitem{KS2001}
K. L. Kuttler and M. Shillor, 
Dynamic bilateral contact with discontinuous friction coefficient, 
{\it Nonlinear Analysis}, 45 (2001), 309--327.

%\bibitem{LIONS}
%J.L. Lions, 
%{\it Quelques m\'ethodes de r\'esolution 
%des probl\'emes aux limites non lin\'eaires}, 
%Dunod, Paris (1969).

\bibitem{MO} 
J. A. C. Martins and J. T. Oden, 
Existence and uniqueness results for dynamic contact problems with 
nonlinear normal and friction interface laws, 
{\it Nonlinear Analysis}, 11 (1987), 407--428.

%\bibitem{MIETTINEN1}
%M. Miettinen,  
%{\it Approximation of Hemivariational Inequalities and 
%Optimal Control Problems,} 
%PhD Thesis, University of Jyv\"askyl\"a, Finland, 
%Report 59 (1993). 

\bibitem{MIGORSKI2} 
S. Mig\'orski, 
On the existence of solutions for parabolic hemivariational inequalities, 
{\it Journal of Computational and Applied Mathematics}, 129 (2001), 77--87. 

\bibitem{MIGORSKI3}
S. Mig\'orski,
Evolution hemivariational inequalities in infinite 
dimension and their control, 
{\it Nonlinear Analysis}, 47 (2001), 101--112.

%\bibitem{MIGORSKI5}
%S. Mig\'orski, 
%Modeling, Analysis and Optimal Control of Systems Governed by 
%Hemivariational Inequalities,
%a chapter in the book {\it "Industrial Mathematics and Statistics"}  
%dedicated to commemorate the Golden Jubilee 
%of Indian Institute of Technology, Kharagpur, India, 2002, 
%J.C. Misra, ed., Narosa Publishing House, 2003.

\bibitem{MIGORSKI6}
S. Mig\'orski, 
Boundary hemivariational inequalities of hyperbolic type 
and applications, {\it J. Global Optimiz.}, 31 (2005), 505--533. 

\bibitem{MIGORSKI7} 
S. Mig\'orski, 
Dynamic hemivariational inequality modeling  
viscoelastic contact problem with normal damped  
response and friction, {\it Applicable Analysis,}  84 (2005), 669--699. 

\bibitem{MIGORSKI8}
S. Mig\'orski and A. Ochal, 
Boundary hemivariational inequality of parabolic type, 
{\it Nonlinear Analysis,} 57 (2004), 579--596. 

\bibitem{MIGORSKIOPT}
S. Mig\'orski and A. Ochal, 
Existence of solutions for second order evolution inclusions 
with application to mechanical contact problems, 
{\it Optimization}, 55 (2006), 101--120. 

\bibitem{MOS1}
S. Mig\'orski, A. Ochal and M. Sofonea, 
Integrodifferential hemivariational inequalities with applications to viscoelastic 
frictional contact, 
{\it Mathematical Models and Methods in Applied Sciences,} 18 (2008), 271--290. 

\bibitem{MOSBOOK}
S. Mig\'orski, A. Ochal and M. Sofonea, 
{\it Nonlinear Inclusions and Hemivariational Inequalities. 
Models and Analysis of Contact Problems}, 
Advances in Mechanics and Mathematics, vol. 26, Springer, New York (2013). 

%\bibitem{MOTP}
%D. Motreanu and P.D. Panagiotopoulos,   
%{\it Minimax Theorems and Qualitative Properties of the Solutions 
%of Hemivariational Inequalities and Applications,} 
%Kluwer Academic Publishers, Boston, Dordrecht, London (1999). 

\bibitem{NP} 
Z. Naniewicz and P. D. Panagiotopoulos,
{\it Mathematical Theory of Hemivariational Inequalities 
and Applications,}  
Marcel Dekker, Inc., New York - Basel - Hong Kong (1995).


%\bibitem{Necas} 
%J. Ne{\v c}as, 
%{\it Les m\'ethodes directes en th\'eorie des 
%\'equations elliptiques,} Masson, Paris (1967).

%\bibitem{NH} 
%J. Ne{\v c}as and I. Hlav\'a{\v c}ek, 
%{\it Mathematical Theory of Elastic and Elasto-Plastic Bodies: 
%An Introduction,} Elsevier, Amsterdam (1981).


%\bibitem{NOWACKI} 
%W. Nowacki, {\it Thermoelasticity,} 2nd edition, Pergamon Press, Oxford (1986).

%\bibitem{OCHAL}
%A. Ochal, 
%{\it Optimal Control of Evolution Hemivariational Inequalities,} 
%PhD Thesis, Jagiellonian University, Cracow, Poland (2001), p.63.

\bibitem{PANA1}
P. D. Panagiotopoulos,  
{\it Inequality Problems in Mechanics and Applications. 
Convex and Nonconvex Energy Functions}, Birkh\"auser, Basel (1985).

%\bibitem{PANA4}
%P. D. Panagiotopoulos, 
%Coercive and semicoercive hemivariational inequalities, 
%{\it Nonlinear Analysis}, 16 (1991), 209--231.

%\bibitem{PANA5}
%P.D. Panagiotopoulos, 
%Modelling of nonconvex nonsmooth energy problems: dynamic hemivariational 
%inequalities with impact effects, 
%{\it J. Comput. Appl. Math.}, 63 (1995), 123--138. 

%\bibitem{PANA6}
%P.D. Panagiotopoulos, 
%Hemivariational inequalities and Fan-variational inequalities. 
%New applications and results, 
%{\it Atti Sem. Mat. Fis. Univ. Modena}, XLIII (1995), 159--191.

\bibitem{PANA7}
P. D. Panagiotopoulos,  
{\it Hemivariational Inequalities, Applications in Mechanics 
and Engineering}, Springer-Verlag, Berlin (1993).

%\bibitem{PP}
%P. D. Panagiotopoulos and G. Pop,  
%On a type of hyperbolic variational-hemivariational
%inequalities, {\it J. Applied Anal.}, 5 (1) (1999), 95--112.

%\bibitem{PPR}
%N.S. Papageorgiou, F. Papalini and F. Renzacci,
%Existence of solutions and periodic solutions for nonlinear
%evolution inclusions,
%{\it Rend. Circolo Mat. di Palermo,} 48 (1999), 341--364.

%\bibitem{Rochdi}
%M. Rochdi, M. Shillor and M. Sofonea,
%A quasistatic contact problem with directional friction and 
%damped response,
%{\it Applicable Analysis,} 68 (1998), 409--422.

\bibitem{RS2000}
M. Rochdi and M. Shillor, 
Existence and uniqueness for a quasistatic frictional bilateral 
contact problem in thermoviscoelasticity, 
{\it Quart. Appl. Math.}, 58 (2000), 543--560.

%\bibitem{RSpreprint}
%M. Rochdi and M. Shillor, 
%A dynamic thermoviscoelastic frictional contact problem 
%with damped response, {\it preprint}, 2001.

%\bibitem{Sofonea}
%M. Sofonea and M. Shillor, 
%A quasistatic viscoelastic contact problem with friction, 
%{\it Comm. Appl. Anal.,} 5 (2001), 135--151. 

%\bibitem{XING}
%G. Xingming, 
%The initial boundary value problem of mixed-typed 
%hemivariational inequality, 
%{\it Intern. J. Math. and Mathematical Sci.}, 
%25 (1) (2001), 43--52. 

\bibitem{ZEIDLER}
E. Zeidler, 
{\it Nonlinear Functional Analysis and Applications II A/B,}
Springer, New York (1990). 

\end{thebibliography}
\end{document}